\documentclass[twocolumn]{aastex631}

\usepackage{hyperref}
\usepackage{amsmath,amstext}
\usepackage[figure,figure*]{hypcap}
\usepackage{amssymb}
\usepackage{nicefrac}
\usepackage{graphicx}
\usepackage{wrapfig}
\usepackage{booktabs}
\usepackage{longtable}
\usepackage{makecell}
\usepackage{comment}

\newcommand\atlas{ATLAS$^{\rm{3D}}$}
\newcommand\romulus{\textsc{Romulus25}}

\shorttitle{Predicting Potential Host Galaxies of Supermassive Black Hole Binaries}

\begin{document}

\title{Predicting Potential Host Galaxies of Supermassive Black Hole Binaries \\ Based on Stellar Kinematics in Archival IFU Surveys}

\correspondingauthor{Patrick Horlaville}
\email{phorlaville24@ubishops.ca}

\author[0009-0007-3541-435X]{Patrick Horlaville}
\affiliation{Department of Physics \& Astronomy, Bishop's University, Sherbrooke, QC J1M 1Z7, Canada}
\affiliation{Trottier Space Institute and Department of Physics, McGill University, 3600 rue University, Montreal, Quebec, H3A 2T8, Canada}

\author[0000-0001-8665-5523]{John J. Ruan}
\affiliation{Department of Physics \& Astronomy, Bishop's University, Sherbrooke, QC J1M 1Z7, Canada}

\author[0000-0002-3719-940X]{Michael~Eracleous}
\affil{Department of Astronomy \& Astrophysics and Institute for Gravitation and the Cosmos, 525 Davey Laboratory, 251 Pollock Road, Penn State University, University Park, PA 16802, USA}

\author[0009-0002-8417-4480]{Jaeden Bardati}
\affiliation{TAPIR, Mailcode 350-17, California Institute of Technology, Pasadena, CA 91125, USA}

\author[0000-0001-8557-2822]{Jessie~C.~Runnoe}
\affil{Department of Physics and Astronomy, Vanderbilt University, Nashville, TN 37235, USA}

\author[0000-0001-6803-2138]{Daryl Haggard}
\affiliation{Trottier Space Institute and Department of Physics, McGill University, 3600 rue University, Montreal, Quebec, H3A 2T8, Canada}

\begin{abstract}

Supermassive black hole binaries (SMBHBs) at the centers of galaxies emit continuous gravitational waves (GWs) at nanohertz frequencies, and ongoing pulsar timing array (PTA) experiments aim to detect the first individual system. Identifying the exact host galaxy of a SMBHB detected in GWs is paramount for a variety of multi-messenger science cases, but it will be challenging due to the large number of candidate galaxies in the sky localization region. Here, we apply recent insights on the distinct characteristics of SMBHB host galaxies to archival galaxy datasets, to predict which nearby massive galaxies are most likely to host SMBHBs detectable by PTAs. Specifically, we use archival galaxy IFU surveys to search for nearby galaxies with distinct stellar kinematic signatures of SMBHB host galaxies, as informed by cosmological simulations. These distinct stellar kinematic signatures, including slow rotation and strong kinematic/photometric misalignments, are a hallmark of recent major galaxy mergers that led to the formation of SMBHBs in these galaxies. We produce a list of nearby massive galaxies that may currently host SMBHBs, ranked by a combination of their host galaxy stellar kinematic properties and their hypothetical GW strain. We discuss how our ranked list can be used (1) for targeted searches for individual sources of continuous GWs by PTAs, (2) to corroborate candidate SMBHBs identified through other approaches, and (3) to select candidate recoiling AGN and closely-separated ($\lesssim$100 pc) dual AGN for telescope follow-up confirmation.

\end{abstract}



\section{Introduction} \label{sec:intro}

In hierarchical galaxy evolution, merging galaxies are predicted to form supermassive black hole binary (SMBHB) systems that emit nHz continuous gravitational waves \citep{Begelman+1980}. As the two supermassive black holes (SMBHs) move through the merged galaxy, they experience dynamical friction, slowing them down through transfer of angular momentum and energy to surrounding stars and dark matter \citep{Chandrasekhar+1943}. This enables them to sink towards the bottom of the gravitational potential well of the merged galaxy \citep{Antonini+2011}, where the two SMBHs will then form a bound binary at $\sim$parsec separation, which hardens through further angular momentum loss from scattering of nearby stars and torques from the surrounding gas \citep{Valtaoja+1989, Quinlan+1996, Armitage+2002, Sesana+2008, Kelley+2017, De_Rosa+2019}. At $\sim$milliparsec separations, the binary system emits continuous gravitational waves (GWs) at nHz frequencies detectable by pulsar timing array (PTA) experiments, before the binary inspirals and coalesces into a single SMBH with a recoil kick \citep{Volonteri+2003, Burke-Spolaor+2019}.

In the coming years, PTA experiments are expected to reach the sensitivity required to detect continuous GWs from individual SMBHB systems at the centers of nearby galaxies. The recent preliminary detection of the stochastic GW background by multiple independent PTA experiments now hints at the existence of an abundant population of SMBHBs at the centers of galaxies in the local Universe \citep[e.g.,][]{EPTA_GWB+2023, NANOGrav_GWB+23, ChinaPTA+2023, ParkesPTA+2023}. As further pulsar timing data accumulate, the detection limit of the GW strain $h_0$ from an individual SMBHB improves. Eventually, PTAs will be able to resolve the localization region of the first individual SMBHB whose GW strain contributes the most to the observed GW background \citep{Sesana+2009, Ravi+2015, Mingarelli+2017, Kelley+2018}.

Identifying the host galaxies of individual SMBHBs detected in GWs from PTAs will be paramount to addressing many key science questions. For example, combining GWs and electromagnetic information originating from the binary environment and its host galaxy will constrain binary orbit parameters, such as the total mass and separation of the binary system \citep{Arzoumanian+2014, Shannon+2015, Lentati+2015, Liu+2021}. These multi-messenger observations will also constrain the interactions that take place in the galaxy nucleus, such as core scouring through three-body interactions with nearby stars \citep{Rajagopal+1995, Jaffe+2003, Wyithe+2003, Enoki+2004, Sesana+2004}. Furthermore, they can constrain cosmological parameters by using SMBHBs as standard sirens \citep{Schutz+1986, Holz+2005}. These science goals will only be realized if the exact host galaxy of the SMBHB detected in GWs can be identified \citep{Bogdanovic+2022}. 

Identifying the host galaxies of individual SMBHBs detected in GWs by PTAs will be challenging. The expected GW sky localization region of the individual SMBHBs detected by PTAs is expected to be of order $10^2$$-$$10^3$ deg$^2$ \citep{Sesana+2010, Goldstein+2018, Truant+2025}. While the mass and distance of SMBHBs detected in continuous GWs are degenerate, they can be determined individually if the gravitational radiation produces a frequency drift \citep{Sesana+2010}. These constraints can be used to make galaxy stellar mass and redshift cuts on the galaxies in the PTA localization region based on empirical galaxy scaling relations. However, the total number of candidate host galaxies even after these selection cuts is still expected to be of order $\sim10^2$ \citep{Goldstein+2019, Petrov+2024, Truant+2025}. Identifying the exact host galaxy among these candidates requires additional approaches.

Many approaches to identifying the host galaxies of SMBHBs are based on periodic variability in the light curves of active galactic nuclei (AGN), but there are currently no secure methods. The light curve variability of AGN can be modulated by a SMBHB, and thus act as an electromagnetic signature for the presence of a SMBHB \citep[e.g.,][]{DOrazio+2015, Charisi+2022, Bogdanovic+2022}. However, such signatures would not be observable if the SMBHB is not actively accreting \citep{Izquierdo+2023, Dong+2023,Truant+2025}, or if the accretion onto the SMBHB is heavily obscured by dust in the nuclear regions \citet{Koss+2018}, which may be a majority of cases. Thus, there is no guarantee that SMBHBs detected in GWs by PTAs will display those light curve variability signatures.

Other studies focused on host galaxies have characterized the population of PTA-detectable SMBHB host galaxies in cosmological simulations, but these results are influenced by galaxy scaling relations. By comparing the properties of the host galaxies of all SMBHBs in the Illustris cosmological simulations to the host galaxies of PTA-detectable SMBHBs, \citet{Cella+2024} found that PTAs are sensitive to SMBHBs in galaxies that are more massive, redder in color, more metal-rich and less star forming. \citet{Saeedzadeh+24} used the \romulus{} cosmological simulations to compare the properties of SMBHB hosts to the overall population of galaxies in the simulation, and found similar results. However, PTAs are only sensitive to the most massive SMBHBs \citep[$M_{\rm{BH}} \gtrsim 10^8 M_\odot$;][]{Sesana+2009, Ellis+2023}. These naturally correspond to more massive, metal-rich, redder, and less star-forming galaxies, due to empirical galaxy scaling relations such as the stellar mass-black hole mass relation \citep[$M_*-M_{\rm{BH}}$;][]{Haring+2004, Torbaniuk+2024}, the stellar mass-metallicity relation \citep[$M_*-Z$;][]{Tremonti+2004, Ma+2016}, the stellar mass-color relationship \citep[$M_*-$color;][]{Law-Smith+2017}, and the stellar mass-specific star formation rate relation \citep[$M_*-$sSFR;][]{Brinchmann+2004, Bauer+2013}. Thus, these properties may not be distinct to SMBHB host galaxies, but instead reflect the sensitivity of PTAs to nHz GWs, which arise from massive SMBHBs in massive galaxies. As a result, characterizing the distinct properties of galaxies that host PTA-detectable SMBHBs beyond their difference in mass relative to the broader galaxy population will require a comparison with a mass-matched control galaxy sample.

Recently, \citet[][hereafter B24A]{bardati+2024a} and \citet[][hereafter B24B]{bardati+2024b} used cosmological simulations of galaxy formation to find that galaxies hosting SMBH mergers and binaries have distinctive morphological and stellar kinematic properties, in comparison to a mass- and redshift-matched control galaxy sample. Their results suggest that the host galaxies of closely separated ($\lesssim$100 pc) SMBH pairs, bound SMBHBs, and recent SMBH mergers tend to have bulge-dominated morphologies in imaging, and have slower rotation with stronger kinematic/photometric misaligned stellar kinematics in integral field spectroscopy. Critically, because their control sample is mass- and redshift-matched to their SMBH merger and binary sample, these characteristics are distinct to SMBHB host galaxies, and do not simply arise from galaxy scaling relations. As such, these results can be used to search for candidate galaxies hosting SMBHBs using archival galaxy datasets, even before PTA detections, effectively making a prediction of which nearby galaxies are likely to host future PTA sources of continuous GWs. 

Other investigations have aimed to predict the host galaxies of individual SMBHBs that will be detected by PTAs among local galaxies by computing the GW strain of hypothetical SMBHBs. \citet{Simon+2014} and \citet{Schutz+2016} computed the GW strain $h_0$ of hypothetical SMBHBs in local galaxies by using the distance and SMBH mass of each galaxy. However, those studies do not incorporate any information on the potential presence of a SMBHB in each galaxy, which is needed to make a more informed prediction.

Here, we predict potential host galaxies of the first SMBHBs that will be detected in GWs by PTAs, by searching archival galaxy datasets for the distinct signatures of SMBH merger and binary host galaxies identified by B24A and B24B. Since most local massive galaxies have already been observed by imaging and integral field spectroscopy surveys, their morphological and stellar kinematic parameters are already available in the literature. We use this information to predict which galaxies among them are the most likely to host SMBHB systems. We also calculate the GW strain $h_0$ of their hypothetical SMBHBs, to identify the potentially strongest individual sources of GWs. Finally, we present a ranked list of galaxies within our sample that both possess the signatures of SMBHB host galaxies and whose hypothetical SMBHBs have the strongest GW strain. These top-ranking galaxies are thus the most likely to be the host galaxy counterpart to individual sources of continuous GWs that will be detected by PTAs in the near future. 

The structure of this paper is as follows. In Section \ref{sec:2}, we determine which set of parameters, from the morphological and stellar kinematic measurements, best discriminates SMBH merger and binary host galaxies against a mass- and redshift-matched sample using the \romulus{} simulations. In Sections \ref{sec:3} and \ref{sec:4}, we describe the archival galaxy datasets we use for our search for SMBHB host galaxies and show the results of our SMBHB host galaxy classification using stellar kinematic features. In Section \ref{sec:5}, we compute the GW strain of the hypothetical SMBHB systems in our sample. In Section \ref{sec:6}, we discuss how our results can be used to predict the host galaxy of the SMBHB that will be detected by PTAs, as well as other applications. We briefly conclude in Section \ref{sec:7}. 

\section{Distinct Signatures of SMBH Merger and Binary Host Galaxies
} \label{sec:2}
\subsection{Background} \label{subsec:2.1}

To identify the distinct morphological and stellar kinematic properties of SMBHB host galaxies in the \romulus{} cosmological simulations, B24A and B24B constructed a sample of SMBH merger and binary host galaxies, as well as a mass- and redshift-matched control sample. In \romulus{}, SMBHs numerically merge in the simulation at a separation of $\sim$700~pc. However, at this stage, the two SMBHs have yet to form a bound binary. Further loss of angular momentum that hardens the binary separation down to $\sim$milliparsec scales, and the eventual physical merger of the SMBHB, are below the resolution limit of the simulation. The time delay between numerical and physical merger is poorly-constrained, and estimates vary widely between 0.1 to 10 Gyr \citep[e.g.,][]{Volonteri+2020, Li+2022}. B24A identified 201 SMBH numerical merger events within the simulation, and tracked their host galaxies up to 1~Gyr after their numerical merger in order to construct a galaxy sample representative of SMBH merger and binary host galaxies. Due to their ignorance of the exact time of the physical SMBHB merger, this sample includes galaxies hosting SMBHBs at separations from $<$700~pc to galaxies hosting SMBHBs that have merged in the past $<$1~Gyr. They also built a control galaxy sample, by selecting galaxies in \romulus{} whose mass and redshift distributions matched those of the SMBH merger and binary host galaxy sample.

After measuring the morphological and stellar kinematic parameters of both their SMBH merger and binary sample as well as their control sample galaxies, B24A and B24B trained a linear discriminant analysis (LDA) predictor to identify the distinct signatures of SMBH merger and binary host galaxies. They first performed stellar population synthesis and radiative transfer simulations to produce synthetic images and stellar kinematic maps of their simulated galaxies, from which they extracted morphological and stellar kinematic parameters. They then trained an LDA predictor to identify the linear combination of parameters that optimally distinguishes the SMBH merger and binary host galaxy sample from the control sample. The resultant LDA predictor assigns a score to each galaxy based on their morphological or stellar kinematic parameters, where high (positive) LDA scoring galaxies are predicted to be more likely to host a SMBH merger or binary. 

The findings of B24A and B24B suggest that SMBH merger and binary host galaxies have distinct morphological and stellar kinematic properties in comparison to a mass- and redshift-matched control sample. Specifically, B24A found that SMBH merger and binary host galaxies are characterized by a more prominent classical bulge in their morphology as probed by imaging, while B24B found that they are characterized by slower rotation and stronger kinematic/photometric misalignments through their stellar kinematics as probed by integral field unit (IFU) spectroscopy. These distinctions are strongest for SMBH merger and binary host galaxies with high chirp mass ($M_{chirp} > 10^{8.2} M_\odot$) and high mass ratio ($q \equiv M_2/M_1 > 0.5$). Those results are consistent with the standard picture of hierarchical galaxy formation, in which major mergers of massive galaxies produce SMBHBs, which result in galaxies with more bulge-dominant morphologies, slower rotation, and more complex stellar kinematics \citep[e.g.,][]{Bois+2011, Naab+2014}. B24A used morphological parameters to derive the LDA predictor:

\begin{equation}\label{eq:LDA_morpho}
    \begin{split}
    \text{LDA} = 1.23 Gini + 0.51M_{20} + 0.52C - 1.04S - 0.01
    \end{split},
\end{equation}

where the Gini coefficient is a measure of how evenly the galaxy flux is distributed, the $M_{20}$ parameter describes the concentration of light in a galaxy, $C$ (concentration) measures the concentration of light in a galaxy relative to its center, and $S$ (smoothness) is a measure of the fraction of light found in clumpy distributions \citep{Lotz+2004, Pawlik+2016}. The LDA predictor from Equation~\ref{eq:LDA_morpho} distinguishes the SMBH merger and binary host galaxies from the mass- and redshift-matched control group with a mean accuracy of $82.6 \pm 3.3\%$. B24B then used stellar kinematic parameters to derive the LDA predictor:

\begin{equation}\label{eq:LDA_kin}
    \begin{split}
    \text{LDA} = 0.51 \log\Delta\mathrm{PA} - 2.81 \lambda_{R_e} + 0.04
    \end{split},
\end{equation}

where $\Delta\mathrm{PA}$ is the difference between the photometric position angle as measured in the galaxy image and the kinematic position angle as measured in the galaxy stellar kinematic map, and $\lambda_{R_e}$ is the spin angular momentum of the galaxy measured at one effective radius $R_e$ \citep{Emsellem+2007}. The LDA predictor from Equation~\ref{eq:LDA_kin} distinguishes the SMBH merger and binary host galaxies from the mass- and redshift-matched control group with a mean accuracy of $85.7 \pm 4.5\%$. Because the LDA equation was constructed by first normalizing the parameters, the absolute value of the coefficients are indicative of each parameter's importance towards the LDA classification. For example, the most important parameter in Equation \ref{eq:LDA_kin} is the $\lambda_{R_e}$ parameter.

\begin{figure}
        \centering
        \includegraphics[width=0.48\textwidth]{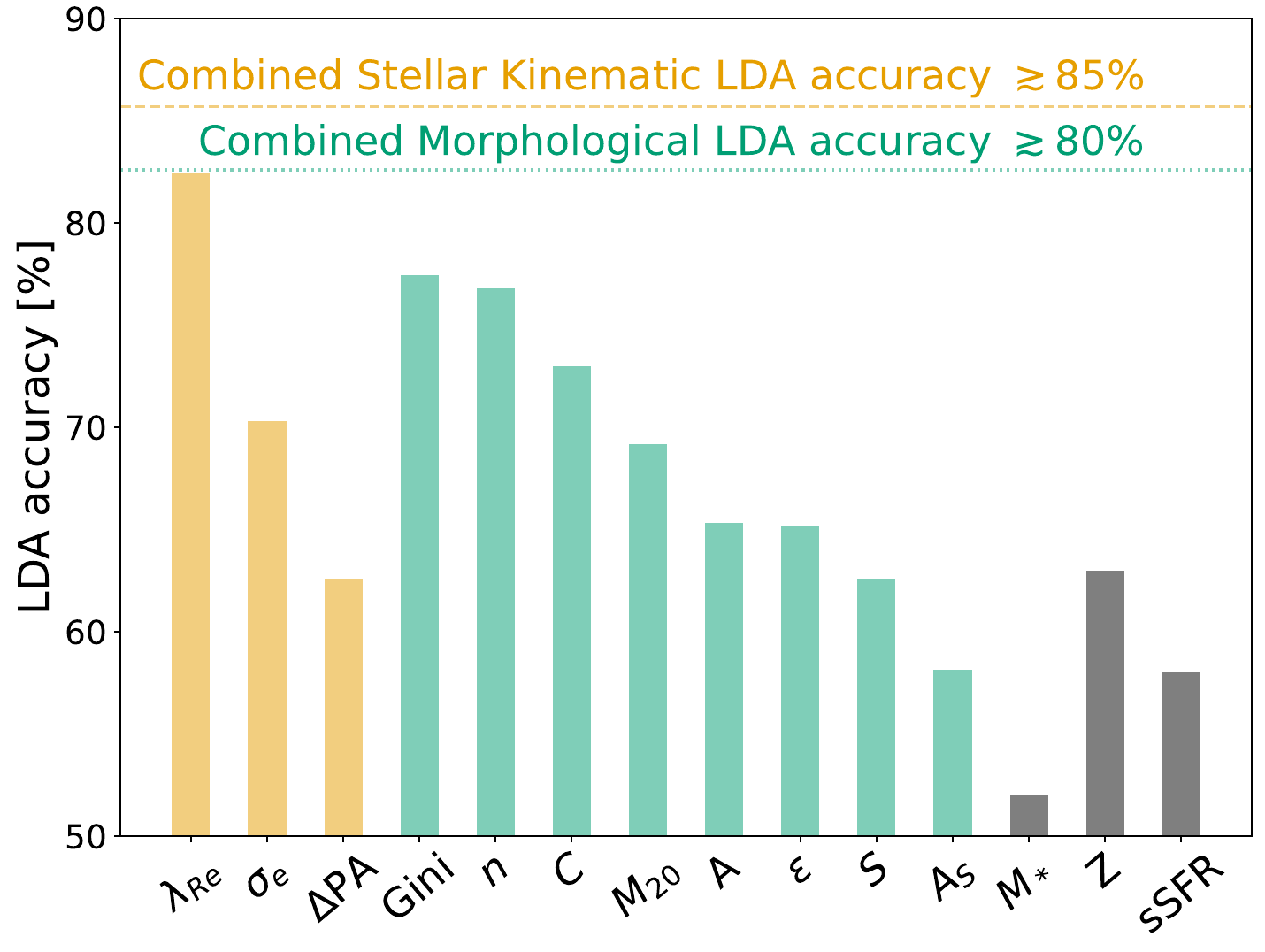}
        \caption{Accuracy of the LDA predictor when trained with individual parameters. The stellar kinematic parameters are indicated with the yellow vertical bars, while the morphological parameters are indicated with the green vertical bars. The dotted green and dashed yellow horizontal lines indicate the accuracies of the full LDA predictors corresponding to Equations \ref{eq:LDA_morpho} and \ref{eq:LDA_kin}, respectively. Additional parameters not listed in Section \ref{subsec:2.1} are the ellipticity $\varepsilon$, the Sérsic index $n$, and the shape asymmetry $A_S$ \citep{Pawlik+2016}. We also show the LDA classification accuracy using the stellar mass $M_*$, the stellar metallicity $Z$, and the specific star formation rate sSFR with the gray vertical bars, which are between $\sim$65\% and $\sim$50\%.}
        \label{fig:LDA_single}
\end{figure}

\subsection{Comparing Stellar Kinematic Signatures to Morphological Signatures}\label{subsec:2.2}

We identify the optimal combination of parameters to search for PTA-detectable SMBHB host galaxies, which will determine which archival datasets we need for our search. In Figure \ref{fig:LDA_single}, we show the accuracy of the LDA predictor, which we train using one parameter at a time while ignoring the others. For the stellar kinematic parameters (yellow bars), the spin angular momentum ($\lambda_{R_e}$) and the stellar velocity dispersion ($\sigma_e$) measured at the effective radius ($R_e$) are the two parameters whose LDA equations are the most accurate at discriminating the SMBH merger and binary host galaxies from the control group. However, when training the LDA using both parameters, the accuracy remains the same as using $\lambda_{R_e}$ alone, because the discriminatory information contained in $\sigma_e$ is degenerate with that from $\lambda_{R_e}$. In contrast, even though $\log \Delta$PA individually has an accuracy of only $\gtrsim$60\%, it contains discriminatory information that is not degenerate with $\lambda_{R_e}$, which causes the accuracy of the LDA trained over both $\lambda_{R_e}$ and $\log \Delta$PA to be higher than $\lambda_{R_e}$ alone. This is why the optimal stellar kinematic equation of the LDA as derived by B24B (Equation \ref{eq:LDA_kin}) contains $\lambda_{R_e}$ and $\log \Delta$PA, but not $\sigma_e$. 

Among all morphological (green bars) and stellar kinematic (yellow bars) parameters shown in Figure~\ref{fig:LDA_single} and present in Equations~\ref{eq:LDA_morpho} and \ref{eq:LDA_kin}, we find that the set of stellar kinematic parameters is the optimal discriminant, motivating the use of archival galaxy IFU surveys. While B24A and B24B separately identified each set of parameters that best classified SMBH merger and binary host galaxies, it is possible that combining both sets of parameters increases the accuracy of the LDA equation. To test this, we train the LDA over the ensemble of both the morphological and stellar kinematic parameters. We use the same approach as B24A and B24B by performing forward stepwise selection to optimally reduce the number of parameters in the LDA equation, through which parameters are added one by one to the LDA. At each step, the corresponding LDA equation is computed to determine whether adding the parameter increases the LDA accuracy or not. This process is repeated until the LDA accuracy decreases. We find that all the morphological parameters as analyzed by B24A are degenerate with either $\lambda_{R_e}$ or $\log \Delta$PA, such that the optimal LDA equation that maximizes the classification accuracy and minimizes the number of parameters is the set of stellar kinematic parameters as derived by B24B, and described in Equation \ref{eq:LDA_kin}. As such, we will mine through archival galaxy IFU surveys to conduct our search of PTA-detectable SMBHB host galaxies.

\section{Archival Galaxy IFU Datasets}\label{sec:3}

\subsection{Sample Selection}

\begin{figure}
        \centering
        \includegraphics[width=0.48\textwidth]{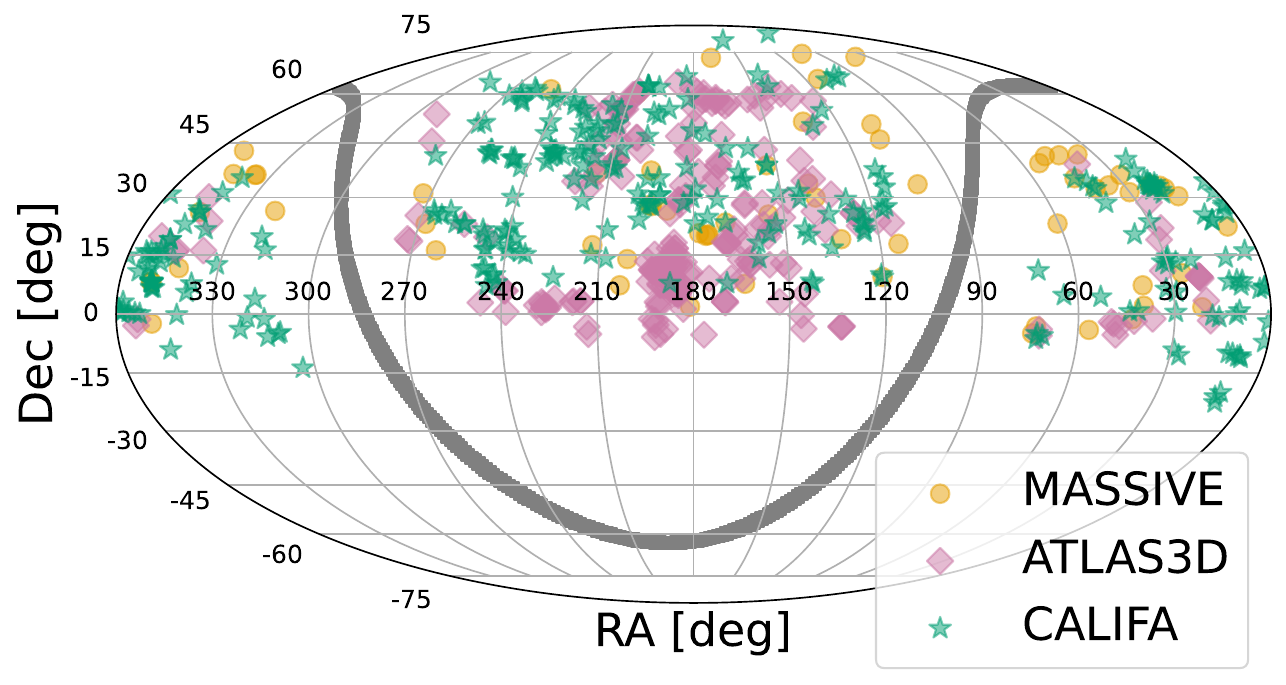}
        \caption{Sky map of the location of the galaxies we use for our search of the potential host galaxies of SMBHBs. The archival galaxy datasets we use (MASSIVE, \atlas{} and CALIFA) cover most of the local massive galaxies in the northern sky. The gray line traces the Galactic plane.}
        \label{fig:skymap}
\end{figure}

We conduct our search for SMBHB host galaxies using archival IFU data from the MASSIVE \citep{Ma+2014}, \atlas{} \citep{Cappellari+2011}, and CALIFA \citep{Sanchez+2012} galaxy surveys.  Since PTAs are sensitive to GWs originating from nearly the full sky, we specifically choose these three IFU surveys because they cover a wide sky footprint. Furthermore, these three surveys are also approximately volume-limited, which mitigates the Malmquist bias that would preferentially select increasingly massive galaxies with increasing distance \citep{Malmquist+1922, Sandage+2000}. While choosing approximately volume-limited surveys is not strictly necessary for our analysis, it allows us to construct a sample of galaxies representative of local massive galaxies. By doing so, we explore all the galaxies in the northern sky within a distance $D < 108$ Mpc and with a stellar mass $M_* \gtrsim 3 \times 10^{11} M_\odot$ for MASSIVE \citep{Ma+2014}, and $D < 42$ Mpc, $M_* \gtrsim 6 \times 10^9 M_\odot$ for \atlas{} \citep{Cappellari+2011}. Although the CALIFA survey is not strictly volume-limited, it has been shown to be $>95\%$ complete for galaxies with stellar mass $5 \times10^{9} M_\odot \lesssim M_* \lesssim 2.5 \times10^{11}M_\odot$ and distance $22~\rm{Mpc}< \textit{D} < 128~\rm{Mpc}$ \citep{Sanchez+2012, Walcher+2014}, so we limit our study to this range. The stellar mass range of the IFU galaxy surveys we use correspond to SMBH mass ranges of $M_\text{BH} \approx 10^{9.6-10.3}~M_\odot$ for MASSIVE,  $10^{7.2-9.8}$$~M_\odot$ for \atlas{} and $10^{7.1-10.0}~ M_\odot$ for CALIFA, where we use the $M_*-M_\text{BH}$ relationship from \citet{Reines+2015} to estimate $M_\text{BH}$. In total, the main MASSIVE, \atlas{} and CALIFA surveys contain 116, 260 and 667 galaxies, respectively. A few galaxies overlap between multiple surveys, and we address how we take this into account in our analysis in Section~\ref{subsec:6.5}. We also note that the redshift ranges of these IFU surveys are different than the redshift range of simulated galaxies used to derive the stellar kinematic signatures of SMBH merger and binary host galaxies, and we discuss why this does not impact our results in Section~\ref{subsec:6.6}. With this selection of galaxies, we cover most of the local massive galaxies in the northern sky that are not in the Galactic plane (see Figure \ref{fig:skymap}).

\subsection{Stellar Mass, Metallicity, and Star-Formation Rate as SMBHB Host Galaxy Discriminants}

\begin{figure}
        \centering
        \includegraphics[width=0.48\textwidth]{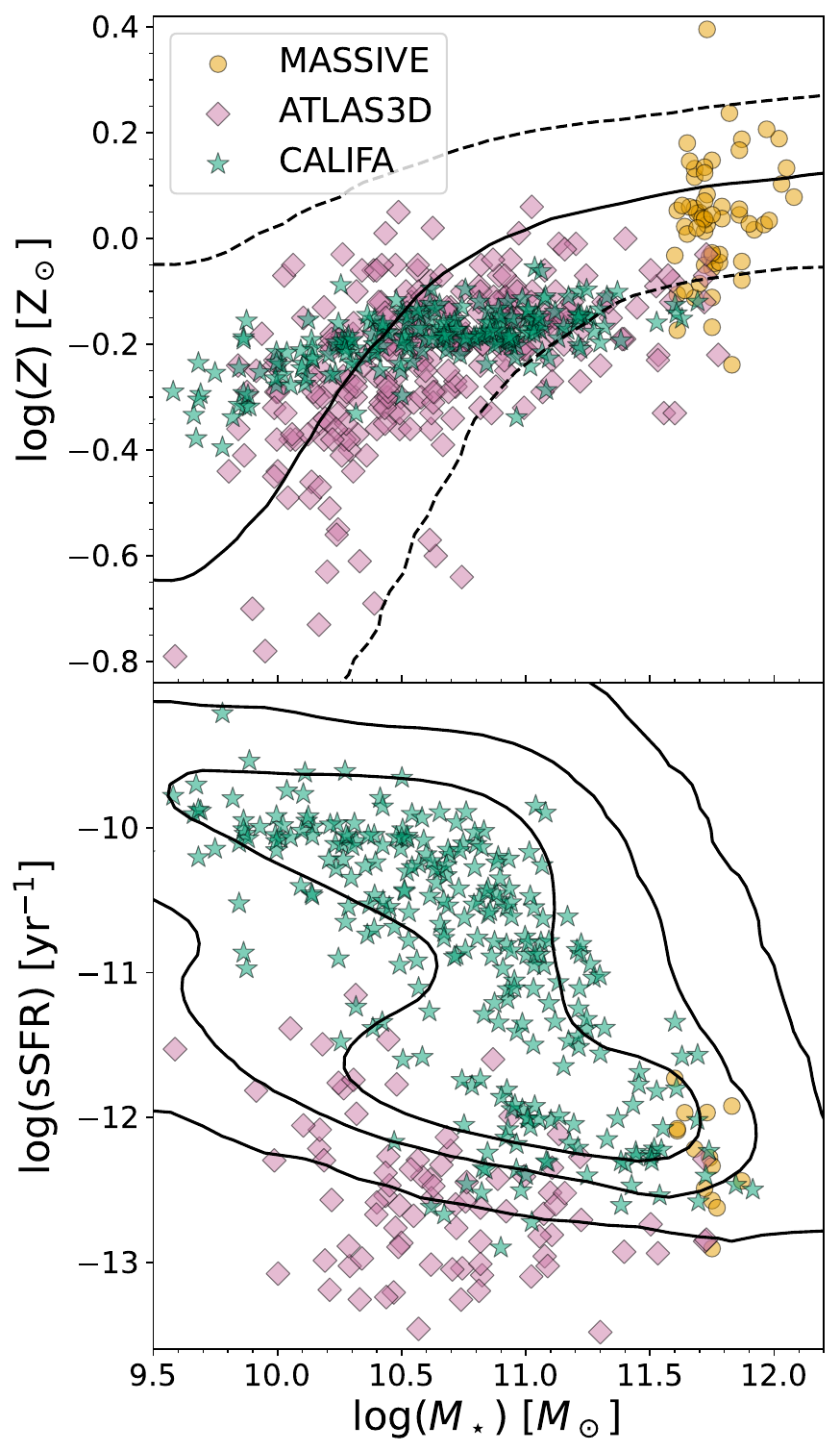}
        \caption{Galaxies within our sample obey well-known global scaling relations. Top panel: stellar mass-metallicity relation ($M_*-Z$) for the MASSIVE, \atlas{}, and CALIFA galaxies for which $\lambda_{R_e}$ and $\Delta$PA are available. The solid black line represents the empirical $M_*-Z$ relation from SDSS galaxies from \citet{Gallazzi+2005}, with the dashed lines representing the $\pm1\sigma$ interval. Bottom panel: stellar mass-sSFR relation ($M_*-$sSFR) for all \atlas{}, MASSIVE and CALIFA galaxies in the top panel for which SFR has also been derived. The contour lines enclose 68\%, 95\%, 99\% of galaxies from the JHU-MPA SDSS galaxy catalog \citep{Brinchmann+2004}.}
        \label{fig:global_corr}
\end{figure}

We first demonstrate that the most massive galaxies within our sample naturally correspond to the most metal-rich and least star-forming, as expected from galaxy scaling relations. As such, identifying the distinct signatures of the host galaxies of SMBHBs detectable by PTAs requires a comparison to a mass-matched control sample of galaxies. Previous studies have asserted from cosmological simulations that PTA-detectable SMBHBs live in galaxies that are more massive, redder in color, more metal rich and less star forming than the overall population of galaxies or the overall population of SMBHB host galaxies \citep{Cella+2024, Saeedzadeh+24}. However, because PTAs are more sensitive to the most massive SMBHBs, we expect their host galaxies to be more massive, which should naturally correlate with higher metallicity and lower star formation rate through galaxy scaling relations. In Figure \ref{fig:global_corr}, we show the $M_*-Z$ and $M_*-$sSFR relations for MASSIVE \citep{Davis+2016, Greene+2019}, \atlas{} \citep{Davis+2014, McDermid+2015}, and CALIFA galaxies \citep{Sanchez+2017, Catalan+2015} for which metallicity and star formation rates have been derived (for a total of 14, 88 and 291 galaxies from MASSIVE, \atlas{} and CALIFA, respectively). We note that the wide range of star formation among the CALIFA galaxies is due to the CALIFA survey probing a different volume in mass and distance than MASSIVE and \atlas{}, thus resulting in galaxies of different morphologies, while \atlas{} and MASSIVE mostly contain E and S0 galaxies, which are usually less star-forming \citep{Gonzalez+2015}. As expected from global mass-scaling relations ($M_*-Z$; \citealt{Tremonti+2004, Ma+2016}, and $M_*$$-$sSFR; \citealt{Brinchmann+2004, Bauer+2013}), Figure \ref{fig:global_corr} reveals a correlation between galaxy stellar mass and metallicity, and an anti-correlation between galaxy stellar mass and specific star formation rate. These correlations are consistent with the paradigm that more massive galaxies tend to have older stellar populations, which through time have enriched the interstellar medium with metals, resulting in higher stellar metallicity, and have exhausted the available gas required to form stars, resulting in lower specific star formation rate \citep[e.g.,][]{Kennicutt+1998, Madau+2014}. Since PTA-detectable SMBHBs are the most massive SMBHBs \citep[with $M_{\rm{BH}} \gtrsim 10^8 M_\odot$;][]{Sesana+2009, Ellis+2023}, they reside in galaxies with high stellar mass. These galaxies are thus naturally more metal-rich and less star-forming compared to the overall population of SMBHB host galaxies or to the overall population of galaxies. These correlations reflect galaxy scaling relations, rather than truly distinctive properties of SMBHB host galaxies. 

Using the \romulus{} simulations, we find that the stellar metallicity and sSFR are not reliable discriminants to identify SMBH merger and binary host galaxies from a mass- and redshift-matched control sample. To assess if $Z$ and sSFR can be used as discriminants of SMBHB host galaxies, we train the LDA equation using $Z$ and sSFR to distinguish the SMBH merger and binary host galaxy sample from the mass- and redshift-matched control galaxy sample in \romulus{}. First, we extract the stellar metallicity and star-formation rate values for each galaxy in our samples from the \romulus{} simulation stellar particle data. We then train the LDA using the metallicity and sSFR parameters individually. We find that the resulting accuracy of the LDA classification is low ($\lesssim65\%$), as shown in Figure~\ref{fig:LDA_single}. We also find that combining both $Z$ and sSFR into the LDA equation does not increase its accuracy. This confirms that PTA-detectable SMBHB host galaxies do not have distinctively-high metallicities and low star formation rates. Rather, they are simply more massive than the overall population of galaxies and SMBHB host galaxies. In contrast, our stellar kinematic LDA predictor in Equation~\ref{eq:LDA_kin} is not affected by galaxy scaling relations, and reflects truly distinctive properties of SMBHB host galaxies. Hence, we compute the LDA score using Equation~\ref{eq:LDA_kin} for galaxies in our archival IFU surveys to identify galaxies most likely to host a SMBHB.

\section{Identifying SMBHB Host Galaxy Candidates in Archival IFU Surveys}\label{sec:4}

\subsection{Retrieval of Stellar Kinematic Parameters}

To identify SMBHB host galaxy candidates from their distinct stellar kinematic properties using Equation~\ref{eq:LDA_kin}, we first retrieve the stellar kinematic parameters of galaxies from archival datasets. Specifically, we retrieve the $\lambda_{R_e}$ and $\Delta$PA parameters from \atlas{} \citep{Emsellem+2011, Krajnovic+2011} and MASSIVE \citep{Veale+2017, Ene+2018}, and $\lambda_{R_e}$ from CALIFA \citep{Falcon+2019}. A few MASSIVE galaxies have no reported value for $\Delta$PA due to not having identifiable kinematic axes which is attributed to the galaxies rotating too slowly \citep{Ene+2018}. Furthermore, only a subset of the full CALIFA sample (galaxies with good quality data and non-disturbed morphologies) have reported values for $\lambda_{R_e}$ \citep{Falcon+2017}. Notably, we do not expect that excluding galaxies with disturbed morphologies might exclude good candidates for PTA-detectable SMBHB host galaxies. Disturbed morphological features associated with galaxy mergers have been shown in cosmological simulations to fade a few hundred Myrs after the numerical merger of the two SMBHs on kpc-scales \citep{DeGraf+2021, bardati+2024a}, which is much quicker than the time required for the two SMBHs to form nHz GW-emitting binary systems detectable by PTAs. With these cuts, the stellar kinematic parameters are available for 71, 260 and 291 galaxies from MASSIVE, \atlas{} and CALIFA, respectively.

Although the CALIFA survey does not provide the $\Delta$PA parameter for its galaxies, we measure $\Delta$PA by computing the morphological position angle PA$\rm{_{morph}}$ and the kinematic position angle PA$\rm{_{kin}}$ from the flux and line-of-sight velocity dispersion (LOSVD) maps, respectively, as produced by the CALIFA collaboration \citep{Falcon+2017}. We retrieve the V1200 (medium resolution) stellar kinematic maps\footnote{\url{https://califa.caha.es/FTP-PUB/dataproducts/Stellar_Kinematics_V1200/}}, and run \texttt{StatMorph} \citep{RodriguezGomez+2019} over the flux and noise maps to compute the morphological position angle $\rm{PA_{morph}}$. We also compute the kinematic position angle $\rm{PA_{kin}}$ by following the technique by \citet{Nevin+2019}, and use the bounded Absolute Radon Transform by \citet{Stark+2018} on the line-of-sight velocity dispersion (LOSVD) maps. This enables us to compute the $\Delta$PA parameter for all 291 CALIFA galaxies in our sample.

\subsection{Selection of Massive Galaxies}

\begin{figure}
        \centering
        \includegraphics[width=0.45\textwidth]{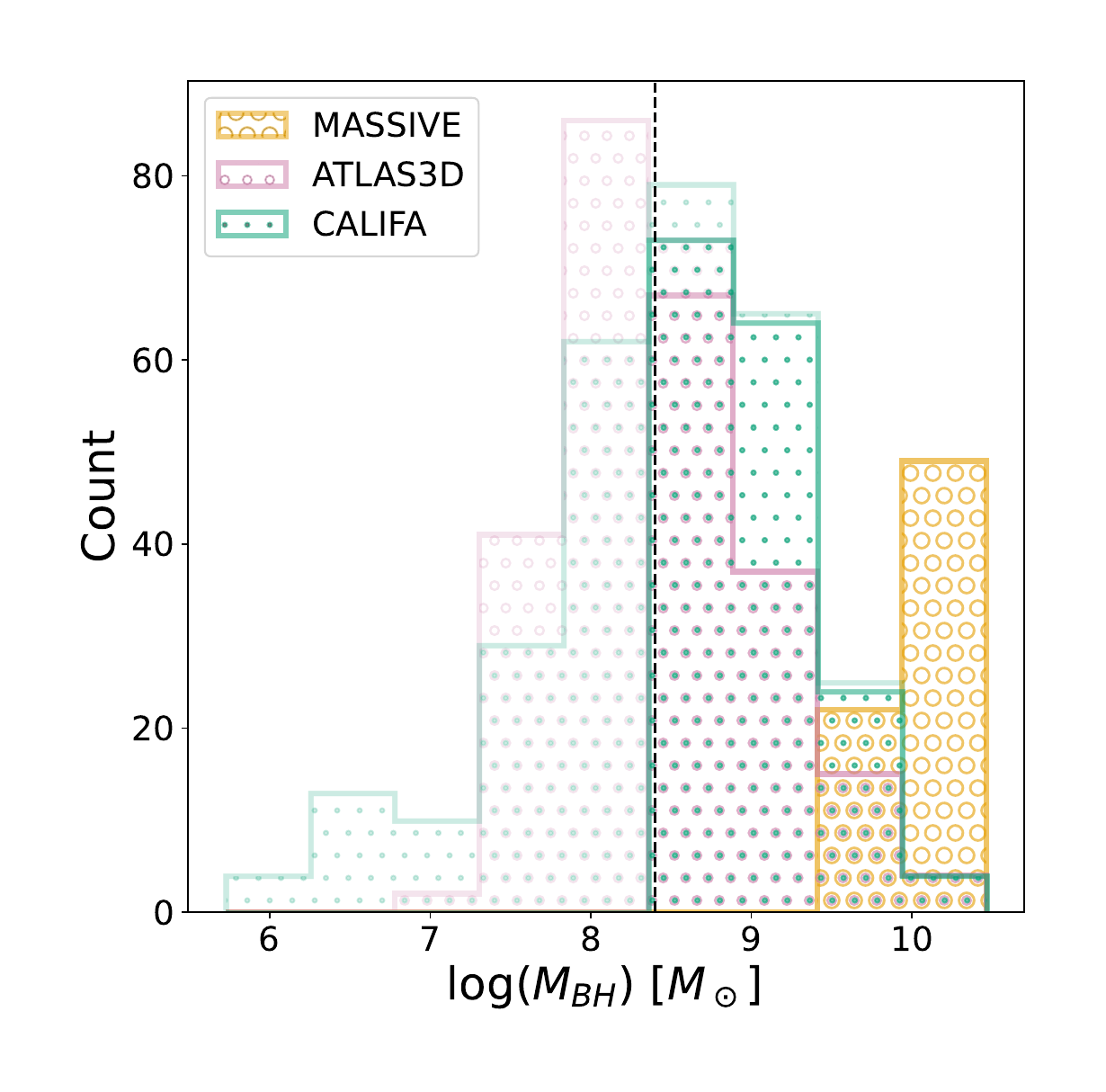}
        \caption{Distribution of the SMBH mass $M_{\rm{BH}}$ of the galaxies in the archival IFU surveys. We search for PTA-detectable SMBHB host galaxies only among galaxies that host the most massive SMBHs ($M_{\rm{BH}} \gtrsim 10^{8.4} M_\odot$, corresponding to $M_{chirp} \gtrsim 10^8 M_\odot$). This minimum SMBH mass threshold is indicated by a black dashed line. The bins with darker lines correspond to galaxies above this threshold.}
        \label{fig:chirp_mass}
\end{figure}

To identify the galaxies in our archival galaxy datasets that are the most likely to host a SMBHB detectable by PTAs, we first select the galaxies whose SMBH mass $M_{\rm{BH}}$ are the highest. PTA experiments are only sensitive to the most massive SMBHB systems, so we select galaxies harboring the most massive SMBHs in the MASSIVE, \atlas{} and CALIFA IFU surveys. We estimate the $M_{\rm{BH}}$ of each galaxy using the empirical $M_* -M_{\rm{BH}}$ relation for elliptical galaxies from \citet{Reines+2015}, and we justify this choice in Section \ref{subsubsec:6.3.1}. The resulting SMBH mass $M_{\rm{BH}}$ distribution of our sample of galaxies is shown in Figure \ref{fig:chirp_mass}.  

We use a fiducial minimum $M_{\rm{BH}} = 10^{8.4} M_\odot$ as the threshold to select the most massive galaxies in our sample. This threshold matches the sensitivity from PTA experiments, and corresponds to a chirp mass $M_{chirp} \sim  10^{8} M_\odot$ for a binary with a mass ratio $q=1$, where $M_{chirp} = [\frac{q}{(1+q)^2}]^{3/5} M_{\rm{BH}}$. Previously, B24B showed that the LDA predictor from Equation~\ref{eq:LDA_kin} has an accuracy of $\gtrsim$85\% in discriminating simulated SMBH merger and binary host galaxies from a mass- and redshift-matched control galaxy sample for SMBHs with a chirp mass $M_{chirp} > 10^{8.2} M_\odot$. Notably, they found that this accuracy does not decrease significantly when the minimum chirp mass threshold is lowered from $10^{8.2} M_\odot$ to $10^{8} M_\odot$. Thus, we still expect the LDA predictor to reach $\sim$85\% accuracy with our adopted minimum SMBH mass threshold of $M_{\rm{BH}}=10^{8.4} M_\odot$, which corresponds to a minimum chirp mass threshold of $M_{chirp} \sim 10^8 M_\odot$. Our selection of the most massive galaxies yields 71, 123 and 165 galaxies in MASSIVE, \atlas{} and CALIFA, respectively.

\subsection{LDA Score and Correlations \\ with $\lambda_{R_e}$, $\Delta$PA, $M_*$, $Z$, and sSFR}

After retrieving the $\lambda_{R_e}$ and $\Delta$PA parameters and selecting the most massive galaxies, we compute the LDA score for each galaxy. We normalize each parameter by subtracting the mean and dividing by the standard deviation of each parameter's distribution, following the method of B24A and B24B, to prevent any one particular parameter from dominating the LDA equation. We then input the normalized parameters in Equation \ref{eq:LDA_kin} to determine the LDA score of each galaxy.

\begin{figure}
        \centering
        \includegraphics[width=0.38\textwidth]{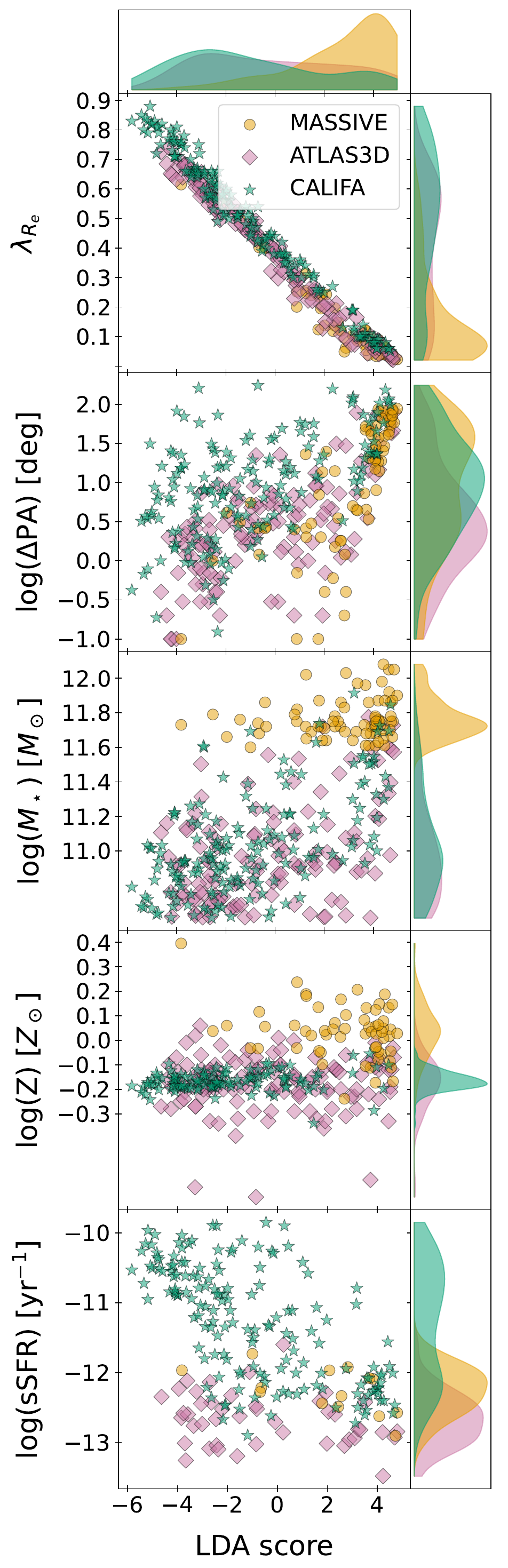}
        \caption{Correlations between various galaxy properties and their LDA score (from top to bottom: $\lambda_{R_e}$, $\log \Delta$PA, stellar mass $M_*$, stellar metallicity $Z$, and specific star formation rate sSFR). Overall, the LDA score has a strong negative correlation with $\lambda_{R_e}$, and a weaker positive correlation with $\log \Delta$PA, but little to no correlation with stellar mass, $Z$, and sSFR.}
        \label{fig:masscut_corr}
\end{figure}

As expected, we find that the LDA score is strongly correlated with $\lambda_{R_e}$, and has a weaker correlation with $\log \Delta$PA. In the first and second panels from the top of Figure \ref{fig:masscut_corr}, we show the correlation between the LDA score and the $\lambda_{R_e}$ and $\Delta$PA parameters for our sample of massive ($M_{\rm{BH}} \gtrsim 10^{8.4} M_\odot)$ galaxies. From the LDA predictor (Equation \ref{eq:LDA_kin}), the absolute values of the coefficients of each parameter are indicative of their relative importance. Thus, it makes sense that the strongest correlation occurs with the $\lambda_{R_e}$ parameter in Figure \ref{fig:masscut_corr}, as its coefficient has an absolute value of 2.81, compared to 0.51 for $\log \Delta$PA. The sign of the coefficients indicates either an anti-correlation (coefficient $<$ 0) or positive correlation (coefficient $>$ 0). This is why $\lambda_{R_e}$ (with a coefficient of $-2.81$) has a strong anti-correlation with the LDA score, while $\log \Delta$PA (with a coefficient of $+0.51$) has a weaker positive correlation with the LDA score. To verify this, we compute Pearson's $r$ \citep[see, e.g.,][]{Bravais+1844, Stigler+1989} for the LDA$-\lambda_{R_e}$ and LDA$-\log \Delta$PA distributions, and find values of $\sim-$1 and $\sim$0.6, respectively, which is consistent with our expectations. We note that the low level of correlation between the LDA score and $\Delta$PA is the same for both higher and lower LDA-scoring galaxies. To verify this, we compute the Pearson's $r$ statistic for both subsets, and find that they remain at $\sim$0.6.

We also find that the LDA predictor does not simply select the most massive galaxies, further confirming that the LDA predictor is identifying the true distinctive stellar kinematic signatures of SMBH merger and binary host galaxies. In the third, fourth, and fifth panels from the top of Figure \ref{fig:masscut_corr}, we show the correlations between the LDA score from Equation \ref{eq:LDA_kin} and galaxy stellar mass, metallicity, and sSFR, respectively. In contrast to $\lambda_{R_e}$ and $\log \Delta$PA, these parameters have little to no correlation with the LDA score. We compute Pearson's $r$ for each of their correlation with the LDA, and find values of $\sim$0.5, $\sim$0.3 and $\sim-$0.3 for $\log M_\star$, $\log Z$ and $\log$ sSFR, respectively. As such, our results show that the LDA predictor is not simply selecting the most massive, metal-rich and least star-forming galaxies, and instead is likely identifying the true distinctive signature of SMBH merger and binary host galaxies.   

\section{Calculating the GW strain of Hypothetical SMBH Binaries}\label{sec:5}

We calculate the GW strain of the hypothetical SMBHB systems in our galaxies. We use the GW strain equation for an equal-mass binary from \cite{Schutz+2016}:

\begin{equation}\label{eq:h0}
\resizebox{.9\hsize}{!}{$h_{0} = 6.9\times10^{-15}  \left( \frac{M_{\rm{BH}}}{10^9 M_\odot} \right)^{5/3} \left( \frac{10~\text{Mpc}}{d_L}\right) \left( \frac{f}{10^{-8} \text{Hz}} \right)^{2/3}$},
\end{equation}

\begin{figure*}
        \centering
        \includegraphics[width=0.66\textwidth]{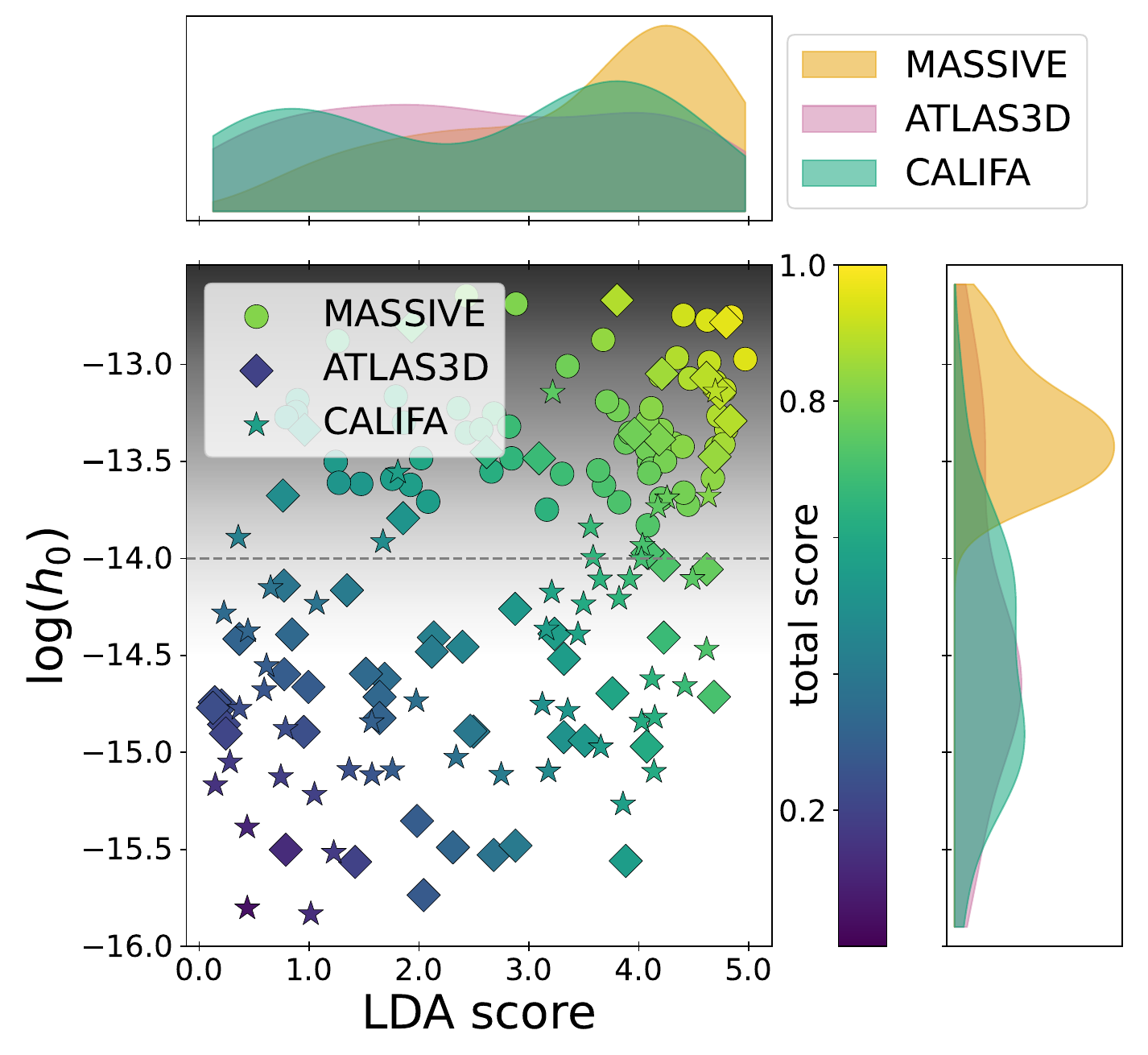}
        \caption{Distribution of LDA score and GW strain for our galaxies in our sample with an LDA score $>$ 0. The color scale represents the total score of each galaxy (Equation~\ref{eq:score}). The gray dashed line represents the approximate $h_0$ sensitivity limit from the 15-year NANOGrav dataset near $f=10$ nHz \citep{Agazie+2024}. Since we calculated the GW strain of the hypothetical SMBHBs within our galaxies using a black hole mass based on their stellar mass, an assumed emitted GW frequency of 10 nHz, and an assumed mass ratio of $q=1$, the dashed line is not a strict but rather an approximate limit, as represented by the gray shaded region around it.}
        \label{fig:total_score}
\end{figure*}

where $h_{0}$ is the GW strain of the hypothetical SMBHB, $M_{\rm{BH}}$ is the total mass of the two black holes ($M_1 + M_2$), $d_L$ is the luminosity distance to the host galaxy, and $f$ is the frequency at which the GWs are assumed to be emitted. Since we have no information about the mass ratio of the potential binary systems, we assume equal-mass SMBHBs (i.e., fiducial mass ratio $q=1$). Since we also do not know the separation between the two black holes, we further assume that the GWs are emitted at a frequency near the highest sensitivity of PTA experiments, around 10 nHz \citep{Arzoumanian+2020}; this corresponds to a separation of $\sim$5 milliparsecs between two black holes of mass $M_{\rm{BH}}=10^{10} M_\odot$ \citep{Schutz+2016}. We discuss the caveats of this hypothetical $h_0$ in Section \ref{subsec:6.2}. 

We combine the GW strain and the LDA score into a total score for each galaxy, based on (1) how likely the galaxy is to host a hypothetical SMBHB, and (2) the amplitude of the GW strain $h_0$ computed with Equation~\ref{eq:h0}. We normalize both the LDA score and the GW strain $h_0$ between 0 and 1, and add them in quadrature to compute a total score. We assign equal weights to each score to remain agnostic about the relative importance between these two metrics towards the likelihood of a PTA detection. We normalize the total score between 0 and 1, such that:

\begin{equation}\label{eq:score}
    \begin{split}
    total~score = \frac{1}{\sqrt 2} \sqrt{\widehat{\rm{LDA}}^2 + \widehat{\log h_0}^2},
    \end{split}
\end{equation}

where $\widehat{\rm{LDA}}$ is the normalized LDA score and $\widehat{\log h_0}$ is the normalized log GW strain score. We compute this total score for each of the massive galaxies in our sample with positive LDA scores. The resulting top ten ranked galaxies are listed in Table \ref{tab:top_rank}, while all ranked galaxies are listed in Table \ref{tab:top_rank_full} in the Appendix. In Figure \ref{fig:total_score}, we present our sample of galaxies with a positive LDA score in the LDA$-h_0$ plane, where the color of each galaxy reflects its total score from Equation \ref{eq:score}. Galaxies with a high LDA score are predicted to be more likely to host a SMBHB, and galaxies whose SMBHBs have a high GW strain $h_0$ are more likely to be detected by PTAs. We discuss some caveats on the interpretation of Figure \ref{fig:total_score} in Section \ref{subsec:6.2}. \\

\begin{deluxetable*}{ccccccccccc}[t!]
\centering
\tablecaption{The top ten highest-ranking galaxies using the total score from Equation \ref{eq:score}. Columns include: galaxy name, total score rank, luminosity distance $D$, IFU survey, log of the black hole mass $M_{\rm{BH}}$, log of the hypothetical GW strain $h_0$, LDA score (Equation~\ref{eq:LDA_kin}), normalized log hypothetical strain $\widehat{\log h_0}$, normalized LDA score $\widehat{\rm{LDA}}$, total score (Equation~\ref{eq:score}), and the inner light profile classification from the literature (where `C'=core, `P'=power-law, and `I'=intermediate). The full list is provided in Table~\ref{tab:top_rank_full} of the Appendix.
\label{tab:top_rank}
}
\tablecolumns{11}
\tablehead{  
\colhead{Name} & \colhead{Total Score} & \colhead{$D$} & \colhead{Survey} & \colhead{$\log M_{\rm{BH}}$} & \colhead{$\log h_0$}  & \colhead{LDA} & \colhead{$\widehat{\log h_0}$}  & \colhead{$\widehat{\rm{LDA}}$} & \colhead{Total} & \colhead{Light} \\ 
\colhead{} & \colhead{Rank} & \colhead{[Mpc]} & \colhead{} & \colhead{$[M_\odot]$} & \colhead{}  & \colhead{Score} & \colhead{} & \colhead{Score} & \colhead{Score} & \colhead{Profile}}
\startdata
\hline
\hline
    NGC4073 & 1 & 91.50 & MASSIVE & 10.06 & -12.76 & 4.84 & 0.97 & 0.97 & 0.97 & C \\
    NGC4486 & 2 & 17.20 & ATLAS3D & 9.61 & -12.78 & 4.80 & 0.96 & 0.96 & 0.96 & C \\
    NGC1060 & 3 & 67.40 & MASSIVE & 9.85 & -12.97 & 4.97 & 0.90 & 1.00 & 0.95 & -- \\
    NGC1016 & 4 & 95.20 & MASSIVE & 10.06 & -12.77 & 4.62 & 0.96 & 0.93 & 0.94 & C \\
    NGC2832 & 5 & 105.20 & MASSIVE & 10.10 & -12.75 & 4.41 & 0.97 & 0.88 & 0.93 & C \\
    NGC0533 & 6 & 77.90 & MASSIVE & 9.88 & -12.99 & 4.64 & 0.89 & 0.93 & 0.91 & -- \\
    NGC7265 & 7 & 82.80 & MASSIVE & 9.81 & -13.13 & 4.78 & 0.85 & 0.96 & 0.91 & -- \\
    NGC0410 & 8 & 71.30 & MASSIVE & 9.79 & -13.09 & 4.70 & 0.86 & 0.94 & 0.90 & -- \\
    NGC4374 & 9 & 18.50 & ATLAS3D & 9.41 & -13.15 & 4.74 & 0.84 & 0.95 & 0.90 & C \\
    NGC4406 & 10 & 16.80 & ATLAS3D & 9.43 & -13.07 & 4.62 & 0.87 & 0.93 & 0.90 & C \\
\enddata
\end{deluxetable*}

\section{Discussion}\label{sec:6}

\subsection{Interpretations of the Ranked List of Galaxies} \label{subsec:6.1}

In our ranked list of galaxies, high-LDA scoring galaxies are the most likely to host SMBHBs, although it is possible that the two SMBHs have either already merged or have yet to form a gravitationally-bound system. As discussed in Section~\ref{sec:2}, galaxies with a high LDA score possess the stellar kinematic signatures of SMBH merger and binary host galaxies, as informed from the \romulus{} cosmological simulations. However, the stellar kinematic signatures identified in \romulus{} (slow rotation and misaligned kinematic/photometric axes) are for SMBHB that have a broad range of separations, from $\lesssim$700~pc (i.e., SMBH pairs) to 0~pc (i.e., recently merged). For observed galaxies in the IFU surveys, the separation of their hypothetical two SMBHs is unknown. Thus, we caution that it is likely that the high-LDA scoring galaxies in our list are contaminated by galaxies harboring SMBHB that have merged in the past Gyr, or have current separations of $\lesssim$100 pc and thus have not yet hardened into a binary system. However, this issue of contamination should not hamper the use of our ranked list of galaxies for targeted continuous GW sources for PTA experiments. Furthermore, the presence of these contaminants in our ranked list enables other science goals, such as searches for close dual AGNs and recoiling AGNs, as we discuss in Section~\ref{ssc:uses} below. We further highlight that the depth and quality of the IFU surveys we use is sufficient for the interpretation of galaxies with the highest LDA scores. The typical relative uncertainty on the stellar kinematic parameters as measured by MASSIVE, \atlas{} and CALIFA is of order $1-10$\% \citep{Emsellem+2011, Veale+2017, Falcon+2019}, which propagates to a similar relative uncertainty on the LDA score. High-scoring LDA galaxies (with LDA scores of $\sim3-4$) conserve a high LDA score under this uncertainty.

We suggest that the host galaxies of gravitationally-unbound SMBH pairs could be distinguished from nHz GW-emitting SMBHB host galaxies based on their inner galaxy surface brightness profile. If the two SMBHs have hardened to a binary system by scouring the galaxy nucleus through three-body interactions with nearby stars, this would result in an observable surface brightness core in the inner light profile of the galaxy \citep[e.g.,][]{Begelman+1980, Hills+1983, Quinlan+1996}. However, other mechanisms such as gravitational wave-induced recoil and tidal deposition have also been proposed to explain the origin of observed surface brightness cores \citep{Nasim+2021, Khonji+2024, Rawlings+2025}, and so this interpretation warrants caution. Nevertheless, we provide the inner surface brightness profile classification of the massive galaxies in our search as found in the literature \citep{Lauer+1995, Faber+1997, Ravindranath+2001, Rest+2001, Laine+2003, Lauer+2005, Lauer+2007a, Krajnovic+2013} in Table~\ref{tab:top_rank} and Table~\ref{tab:top_rank_full}, if one desires to interpret it as a discriminant between potential SMBH pair and SMBHB host galaxies. 

\subsection{Use Cases of the Ranked List of Galaxies}
\label{ssc:uses}

The most basic use of our ranked list of galaxies in Tables~ \ref{tab:top_rank} and \ref{tab:top_rank_full} is to identify the most likely host galaxy counterpart to an individual PTA source. Once PTA experiments identify an individual source of continuous GWs, nearby massive galaxies in its sky localization region that are also highest-ranked in our list are the most likely counterparts. However, even before such a PTA detection, our ranked list of galaxies can be used for several science cases, as discussed below.

\subsubsection{Targeted Searches for Individual Continuous GW Sources by PTAs}

The highest-ranked galaxies in our ranked list can already be used in targeted searches for individual continuous GW sources in PTA data. PTA searches for continuous GWs from an individual SMBHB typically fit pulsar timing residual data to models with many parameters, such as the sky position, the GW frequency $f$, the orbital phase, the GW polarization angle, the orbital inclination, and others, depending on the chosen model \citep[e.g.,][]{Arzoumanian+2020, Liu+2021}. These searches are computationally expensive, but can be sped up by fixing one or more of these parameters. For example, targeted searches fix the location in the sky of the GW source to observed galaxies, which improves the detection sensitivity \citep[e.g.,][]{Liu+2023, Charisi+2024}. Our high-ranking galaxies are among the best PTA-detectable SMBHB host galaxy candidates in the northern sky, and thus targeted PTA searches can focus on them for faster and more sensitive searches.

\subsubsection{Independent Corroboration of Candidate SMBHBs Discovered Through Other Means}

Our ranked list of galaxies can be used to independently corroborate candidate SMBHBs discovered through other means (e.g., light curve periodicities), based on the host galaxy stellar kinematic properties. Our full galaxy list in Table~\ref{tab:top_rank_full} provides the stellar kinematic properties of the overall population of massive nearby galaxies, spanning those likely to host SMBHBs (high LDA score) to those unlikely (low LDA score). For a candidate SMBHB discovered through other means, a comparison of its host galaxy stellar kinematics to our full galaxy list can provide independent evidence to either reinforce or weaken the hypothesis that its central SMBH is indeed a SMBHB. Specifically, the stellar kinematic parameters $\lambda_{R_e}$ and $\Delta$PA of the candidate SMBHB's host galaxy can be measured from IFU spectroscopy, and used to compute a LDA score using Equation~\ref{eq:LDA_kin}. The resultant LDA score can then be compared to our full galaxy list in Table~\ref{tab:top_rank_full}; if the SMBHB candidate has a high LDA score compared to the galaxies in our full list, this corroborates the hypothesis that the galaxy hosts a SMBHB, and vice versa. We emphasize that since this comparison does not involve the GW strain $h_0$, it is the LDA score (Equation~\ref{eq:LDA_kin}) in Table~\ref{tab:top_rank_full} that should be considered, rather than the total score (Equation~\ref{eq:score}). We also note that the LDA score in Equation~\ref{eq:LDA_kin} requires normalized values of $\lambda_{R_e}$ and $\Delta$PA for the candidate SMBHB's host galaxy. To perform this normalization, one should subtract the mean value of the parameter from our distribution and divide by its standard deviation (the mean and standard deviation of $\lambda_{R_e}$ and $\log\Delta$PA from our full list in Table~\ref{tab:top_rank_full} are $(\mu_{\lambda_{R_e}}, \sigma_{\lambda_{R_e}})= ({0.39, 0.25})$ and $(\mu_{\log\Delta\mathrm{PA}}, \sigma_{\log\Delta\mathrm{PA}})=({0.78}, 0.73)$, respectively). The resulting normalized values of $\lambda_{R_e}$ and $\Delta$PA for the candidate SMBHB's host galaxy can then be directly inserted into Equation~\ref{eq:LDA_kin}. 

\subsubsection{Identifying Candidate Dual AGNs and Recoiling AGNs for Follow-Up Observations}

Our list of galaxies can also be used to select candidate closely separated ($\lesssim$100 pc) dual AGN, which can be confirmed with follow-up observations. As discussed in Section~\ref{subsec:6.1}, some of our top-ranked galaxies may instead host close ($\lesssim$100 pc) SMBH pairs, since their host galaxy stellar kinematics are similar to SMBHBs and thus contaminate our ranked list. If both SMBHs in the pair are accreting, then they could be observable as a close dual AGN. Follow-up telescope imaging of these candidates with high spatial resolution in the infrared, X-ray, or radio could spatially resolve the two AGNs for confirmation. Furthermore, if the two SMBHs have not yet formed a bound $\sim$parsec-scale binary (and are thus potentially resolvable in follow-up observations), the inner surface brightness profiles of the host galaxy would not have been modified by scouring yet. In this scenario, the host galaxies would thus display a power-law inner surface brightness profile (rather than a core), which can be used as an additional selection cut to maximize the efficiency of target selection for follow-up observations. Within our sample, {ten} galaxies (NGC 6703,  NGC 3414, NGC 474, NGC 4494, NGC 3226, NGC 4596{, NGC 5831, NGC 4621, NGC 2859 and NGC 4442}) have a positive LDA score and a power-law inner surface brightness profile.

In addition to dual AGN, we suggest that our ranked list of galaxies can be used to select candidate recoiling AGN that can be verified with follow-up observations. As discussed in Section~\ref{subsec:6.1}, our top-ranked galaxies are also likely to host SMBHBs that have merged in the past $<$Gyr. For a merging SMBHB, anisotropic emission of GWs produce a recoil kick on the merged SMBH \citep[e.g.,][]{Campanelli+2007, Schnittman+2007, Blecha+2011}, with velocities ranging from $\lesssim$500 km/s to 4000 km/s depending on the binary parameters \citep{Bogdanovic+2007, Campanelli+2007}. If the kicked SMBH is accreting, it could be observed as an AGN with spatial or kinematic offsets from the host galaxy. Such methods have already been employed towards detecting AGN recoil candidates, for example using HST images \citep[e.g.,][]{Lena+2014}. In contrast with the dual AGN discussed above, the inner light profile of the host galaxy of a recently-merged SMBHB would be scoured and display a core. Follow-up observations of our top-ranked galaxies, especially those with cores, could reveal these spatial or kinematic offsets  \citep{Blecha+2011, Kim+2016}. 

\subsection{Uncertainties in Calculating the GW Strain $h_0$} \label{subsec:6.2}

\subsubsection{Black Hole Mass}\label{subsubsec:6.3.1}

Our calculation of the SMBH mass $M_{\rm{BH}}$ carries a statistical error stemming from our use of an empirical scaling relation, which causes the GW strain $h_0$ as shown in Figure~\ref{fig:total_score} to be approximate. Specifically, the empirical $M_* - M_{\rm{BH}}$ relation we use from \cite{Reines+2015} has a scatter of $\sim$0.5 dex. Since $h_0 \propto M_{chirp}^{5/3} \propto M_{\rm{BH}}^{5/3}$, this would lead to a scatter of a factor of $\sim$7 in $h_0$, which corresponds to an uncertainty of $\sim$0.8 dex on the y-axis of Figure~\ref{fig:total_score}. Among our top ranking galaxies in Table~\ref{tab:top_rank}, two (NGC 4486 (M87) and NGC 4374 (M84)) have dynamically-or directly-measured SMBH masses in the literature. The SMBH mass of M87 has been directly measured from the black hole's shadow by \citet{EHT+2019} to be $M_{\rm{BH}}\sim 6.5 \times 10^{9} M_\odot$, within 1$\sigma$ of our value found with stellar mass. The SMBH mass of M84 has been measured through gas kinematics by \citet{Bower+1998} to be $M_{\rm{BH}}\sim 1.5 \times 10^{9} M_\odot$, also within 1$\sigma$ of our stellar mass value. SMBH masses of other galaxies such as NGC 2832 and NGC 4874 have also been published, although they were found through the $M_{\rm{BH}}-\sigma_e$ empirical relation \citep{Schutz+2016, Dullo+2019}. Their values are below our adopted estimates using a $M_*-M_{\rm{BH}}$ relation (a difference of up to 1.5 dex), which is consistent with recent studies suggesting that the $M_{\rm{BH}}-\sigma_e$ relation systematically underestimates $M_{\rm{BH}}$ for the most massive galaxies, and is therefore a less robust method to determine $M_{\rm{BH}}$ of local massive galaxies \citep[e.g.,][]{Lauer+2007b, Dullo+2021, Liepold+2024}. This further justifies our choice of the $M_*-M_{\rm{BH}}$ relation to compute $M_{\rm{BH}}$. 

\subsubsection{Black Hole Mass Ratio}

Our assumption of an equal mass ratio $q=1$ in the calculation of the chirp mass $M_{chirp}$ also leads to a systematic error in the gravitational wave strain $h_0$. Although the mass ratio of hypothetical SMBHBs in our galaxies is unknown, our choice of $q=1$ may be justified because accretion onto the binary would drive it towards equal mass \citep{Young+2015}. However, hypothetical SMBHBs in our galaxy dataset might have a mass ratio $0 < q \leq 1$. Since $h_0 \propto M_{chirp}^{5/3} \propto \frac{q}{(1+q)^2}$, if we assume a SMBHB to have a mass ratio of $q=1$ while its actual mass ratio is, e.g., $q=0.1$, we would overestimate its GW strain $h_0$ by a factor of $\sim$3, which corresponds to a $\sim$0.5 dex difference on the y-axis of Figure \ref{fig:total_score} (but still smaller than the uncertainty caused by the black hole mass). Therefore, the GW strain $h_0$ of galaxies shown in Figure~\ref{fig:total_score} is an upper limit. 

\subsubsection{Binary Separation} \label{subsubsection:6.3.3}
    
Ultimately, our lack of constraints on the hypothetical SMBHB orbital separation is our most significant obstacle towards identifying potential PTA-detectable SMBHB candidate host galaxies. In Figure \ref{fig:total_score}, we assume a GW emission frequency of 10 nHz to compute the GW strain $h_0$, which corresponds to a binary separation of $\sim$5 milliparsecs for a system of mass $M_{\rm{BH}}\sim10^{10}M_\odot$ \citep{Schutz+2016}. As discussed in Section~\ref{subsec:6.1}, the exact separations can be anywhere between $\sim$100~pc to 0~pc (already merged). PTA-detectable SMBHBs need separations of $\sim$milliparsec to emit nHz GWs, which we cannot constrain (although they would rank highly in our list if there are any in our archival galaxy datasets). Thus, the galaxies shown in Figure \ref{fig:total_score} that are above the current NANOGrav $h_0$ sensitivity limit are not necessarily predicted to host SMBHBs currently detectable by PTAs, but may instead host SMBHB systems that have either already merged in the past Gyr, or that are currently at larger separations and have not yet hardened into a bound nHz-emitting SMBHB.

\subsection{Cross-Referencing with Multi-AGN Catalogs}

Through cross-referencing with multi-AGN catalogs, we find five candidate recoiling AGN host galaxies in our galaxy dataset, all of which have a positive LDA score, which is consistent with our interpretation. We use the \citet{Pfeifle+2024} Big Multi-AGN Catalog, which is a census of all known and candidate multiple-AGN systems in the literature, and identify the galaxies in our sample that are present in that catalog. Notably, we find that five of our \atlas{} galaxies host candidate recoiling AGN \citep[NGC 4486 (M87), NGC 4168, NGC 4278, NGC 4636 and NGC 5846, as identified by][]{Lena+2014}. These five galaxies all have $\lesssim$10 pc spatial offsets between the AGN and the galaxy center, which suggests recent ($\lesssim$0.1 Myr) SMBH mergers assuming typical recoil velocities of $\sim$100--1000 km~s$^{-1}$ \citep{Campanelli+2007}. We also find one galaxy (NGC 6338) from the CALIFA sample in the \citet{Pfeifle+2024} catalog as identified by \citet{Comerford+2014} that hosts a candidate dual AGN, with a projected separation of 1.6~kpc. All six of these galaxies have positive LDA scores, which is consistent with our expectation that the LDA score not only identifies SMBHB host galaxies, but also the host galaxies of SMBH pairs and recent SMBHB mergers.

Other observational evidence also suggests that many of the highest-ranked galaxies in our sample either have undergone a recent galaxy merger, host a SMBHB, or host a recent SMBH merger. In particular, our \#1 scoring galaxy, NGC 4073, has been observed by \citet{Lauer+2005} to display a local minimum in its surface brightness profile near its center. Surface brightness central minima have been hypothesized to result from the hardening and recent merger of a SMBHB \citep{Holley-Bockelmann+2000, Lauer+2002}, which is consistent with expectations from our LDA score. Our \#6 scoring galaxy, NGC 533, lies in a galaxy group that is suspected to have recently undergone a merger based on X-ray observations of its intragroup medium gas \citep{Finoguenov+2007, Gu+2012}. We also note that the majority (6/10) of our highest-ranked galaxies reside in either galaxy groups or low-mass clusters rather than massive galaxy clusters, based on cross-referencing with galaxy catalogs \citep{Mahtessian+1998}. This is consistent with the scenario in which galaxy groups are more conducive to major mergers of massive galaxies that lead to the formation of SMBHBs, due to the lower velocity dispersions of galaxies in the group \citep{Binney+2008}.

\subsection{Discrepancies Among Overlapping Galaxies Between Different Surveys} \label{subsec:6.5}

Among the 71, {260 and 291} galaxies from the MASSIVE, \atlas{}, and CALIFA surveys that we use to search for PTA-detectable SMBHBs, there are 18 galaxies that overlap between two different galaxy surveys, and none that are in all three. In particular, NGC 5353, NGC 4472, NGC 5557, and NGC 5322 are present in both MASSIVE and \atlas{}, NGC 2592, NGC 5631, NGC 2880, NGC 5485, and NGC 6278 are present in both \atlas{} and CALIFA, while NGC 7619, NGC 4816, NGC 1167, NGC 3615, NGC 3158, NGC 2513, NGC 0499, NGC 1060, and NGC 4874 are present in both MASSIVE and CALIFA. As such, many of these galaxies have different reported distances and stellar kinematic parameters, depending on the survey.

For overlapping galaxies, we find that the stellar kinematic parameters are similar between the different surveys within the uncertainties, whereas some galaxy distances have been measured differently and disagree. For the stellar kinematic parameters, we find a difference of about $1 \pm 5$\% between the reported values of the stellar angular momentum $\lambda_\mathrm{Re}$ of overlapping galaxies, which is within the $\sim1-10$\% typical uncertainty for $\lambda_\mathrm{Re}$ reported by the IFU surveys we use \citep{Emsellem+2011, Veale+2017, Falcon+2019}. For the galaxy distances, we use those that have been corrected with the surface brightness fluctuation method or corrected for local peculiar velocity, whenever available. For example, NGC 1060 as observed by MASSIVE is ranked \#5 in our list, but \#{13} as observed by CALIFA. Upon closer inspection, this difference arises not from a different LDA score (both are similar), but rather from a difference in GW strain $h_0$ due to MASSIVE reporting a distance of 67.4~Mpc and CALIFA reporting a distance of 73.9~Mpc. Since the MASSIVE galaxy distance is smaller, its $h_0$ is higher and its rank is higher. In this case, the distance of NGC 1060 from MASSIVE has been corrected for its local peculiar velocity, and thus we use the distance reported in MASSIVE.

\subsection{Redshift Difference Between \\ the Simulated and Observational Galaxy Datasets} \label{subsec:6.6}

Although the redshift range of the IFU surveys we use is lower than the redshift range of the \romulus{} simulated galaxies, this does not significantly affect our results, because the distinct stellar kinematic signatures of simulated SMBH merger and binary host galaxies do not display redshift evolution. The simulated galaxies in \romulus{} used by B24B to derive the LDA predictor have redshifts of $0.5 \lesssim z \lesssim 2$ (mean of $z\sim1.5$). As such, it is possible that the stellar kinematic signatures of SMBH merger and binary host galaxies identified by B24B are different for lower redshift galaxies, such as those at $z\lesssim0.03$ in the IFU surveys we use. Although the \romulus{} simulation does not contain enough SMBH merger and binary host galaxies to directly assess if the LDA predictor in Equation~\ref{eq:LDA_kin} applies at $z\sim0$, we test whether this LDA predictor evolves over redshift. Specifically, we retrain the LDA predictor only for a subset of simulated galaxies around redshift $z\sim 0.5$, and find that the resulting LDA equation is similar to Equation~\ref{eq:LDA_kin} in terms of parameters and parameter coefficients. This explicitly demonstrates that the distinct stellar kinematic properties of simulated SMBH merger and binary host galaxies do not evolve with redshift, and should apply even to galaxies at lower redshifts.

Despite the different redshift ranges of the simulated and observed galaxies, the synthetic IFU data used by B24B to compute the LDA predictor probe similar physical scales in the galaxies compared to the observational IFU datasets, which enables us to directly use the LDA predictor on our archival galaxy datasets. As a consequence of the difference in redshift range between the simulated and observed galaxies, if the physical scale probed by the synthetic IFU data for the simulated galaxies used to compute the LDA predictor is much smaller than the physical scale probed by the observational dataset, the LDA predictor may not be directly applicable to the observational IFU dataset. In particular, the pixel spatial resolution of the synthetic stellar kinematic maps produced by B24B probes physical scales of 400 to 900~pc in each galaxy, depending on the redshift of the simulated galaxy. On the other hand, the physical scales probed by the stellar kinematic maps from CALIFA, \atlas{}, and MASSIVE range from $\sim$300--1200~pc, depending on redshift and instrument resolution. Thus, despite the difference in redshift ranges, the physical scales probed are similar, enabling us to directly apply the LDA predictor in Equation~\ref{eq:LDA_kin} to the MASSIVE, \atlas{} and CALIFA galaxy IFU datasets.

\section{Conclusions}\label{sec:7}

We mine archival galaxy IFU surveys to search for potential candidate host galaxies of individual SMBHBs that could be detected in GWs by PTAs. To do this, we use results from the \romulus{} cosmological simulations to identify the optimal set of parameters that discriminate simulated SMBH merger and binary host galaxies from a mass- and redshift-matched control sample. This selection is embedded in a LDA predictor. We then compute the LDA score for galaxies from archival IFU surveys to identify nearby galaxies that display the distinct signatures of SMBH binary and merger host galaxies. Assuming that their hypothetical two SMBHs are equal mass and are at $\sim$milliparsec separations, we calculate their expected gravitational wave strain $h_0$. We combine the LDA score with $h_0$ to calculate a total score for each galaxy that reflects both (1) how likely they are to host a SMBHB, and (2) the strength of their hypothetical gravitational wave strain. Our main findings are as follows:
    
\begin{enumerate}
   \item Using the \romulus{} cosmological simulations, we determine that among the full set of morphological and stellar kinematic parameters, it is the set of stellar kinematic parameters that optimally discriminates SMBH merger and binary host galaxies from a mass- and redshift-matched control sample. By selecting simulated galaxies with chirp mass $M_{\rm{chirp}}>10^{8} M_\odot$ and mass ratio $q>0.5$, the accuracy of this classification reaches $\gtrsim$85\%.
   
   \item We derive the stellar kinematic signatures of simulated SMBH merger and binary host galaxies using a mass- and redshift-matched control sample, such that our results are not affected by galaxy scaling relations. We apply these distinctive stellar kinematic signatures (slower rotation and stronger kinematic/photometric misalignments) to archival IFU observations of massive nearby galaxies, to predict which ones are the most likely to host a SMBH merger or binary.
   
   \item We produce a ranked list of galaxies that correspond to the best candidates for the host galaxies of individual SMBHBs in the northern sky that will be detected in GWs by PTAs. Even before a PTA detection, this list can be use to (1) perform targeted searches for individual sources of continuous GW in PTA data, (2) to corroborate candidate SMBHBs discovered through other means, and (3) to select candidate closely-separated ($\lesssim$100 pc) dual AGNs and recoiling AGN for follow-up observations. 
   
\end{enumerate}

The quest to detect continuous nHz GWs from an individual SMBHB is ongoing through multiple PTA experiments, and their detections will require both observational and theoretical progress. While we focus in this paper on identifying potential PTA GW sources among massive nearby galaxies using archival IFU surveys in the northern sky, ongoing galaxy IFU surveys in the south such as the Hector Galaxy Survey\footnote{https://hector.survey.org.au/} can extend these efforts for full sky coverage of individual PTA sources. Such a search may become especially pressing given the recent tentative detection of a GW hotspot by the MeerKAT PTA collaboration at southern declinations not covered here \citep{Grunthal+2024}. On the theoretical front, future studies could attempt to characterize more distinctly the host galaxies of nHz GW-emitting SMBHBs, e.g. by resolving the gravitational dynamics of the SMBHs \citep[with, e.g., KETJU;][]{Mannerkoski+2023} in large cosmological simulations and by performing more high-resolution zoom-in simulations of SMBHB host galaxies \citep[e.g.,][]{Volonteri+2020}, or by discovering new observable signatures of SMBHB host galaxies (related to, e.g., gas kinematics). When the host galaxy of an individual SMBHB detected through GWs by PTAs is identified, telescope follow-up across the electromagnetic spectrum will provide insights on the formation and environments of SMBHBs that remain mysterious. 

\begin{acknowledgments}

J.J.R.\ and J.C.R.\ thank the organizers and attendees of The Era of Binary Supermassive Black Holes conference at the Aspen Center for Physics for insightful discussions. The authors also thank the anonymous referee for their relevant feedback, which enhanced the quality of the present manuscript.

P.H.\ acknowledges support from a NSERC Canada Graduate Scholarship, a Fonds de recherche du Québec -- Nature et technologie (FRQNT) MSc Scholarship, and a Graduate Entrance Scholarship from the Bishop's University Foundation. Bishop’s University is located on the traditional territory of the Abenaki people (the people of the rising sun). We acknowledge their stewardship and appreciate our status as guests on Abenaki territory. J.J.R.\ acknowledges support from the Canada Research Chairs (CRC) program, the Natural Sciences and Engineering Research Council of Canada (NSERC) Discovery Grant program, the Canada Foundation for Innovation (CFI), and the Qu\'{e}bec Ministère de l’\'{E}conomie et de l’Innovation. J.C.R.\ acknowledges support from the National Science Foundation (NSF) from grant NSF AST-2205719, and the NASA Preparatory Science program under award 20-LPS20-0013. D.H.\ acknowledges funding from the NSERC Arthur B. McDonald Fellowship and Discovery Grant programs, and the CRC program. The authors acknowledge support from the Centre de recherche en astrophysique du Québec, un regroupement stratégique du FRQNT.
 
 \end{acknowledgments}

\vspace{5mm}
%\facilities{HST (WFPC2)}


\software{\texttt{astropy} \citep{astropy18}; \texttt{Matplotlib} \citep{Matplotlib+2007};
\texttt{SciPy} \citep{SciPy+2020};
\texttt{NumPy} \citep{Numpy+2020};
\texttt{StatMorph} \citep{RodriguezGomez+2019}; \texttt{scikit-learn} \citep{ScikitLearn+2011}; \texttt{PyEphem} \citep{PyEphem+2011}}



\pagebreak

\appendix
\onecolumngrid

\section{Full Ranked List}\label{sec:appendix}

\scriptsize
\renewcommand{\arraystretch}{1}
\setlength{\LTleft}{0pt}
\setlength{\LTright}{0pt}
\setlength{\tabcolsep}{3pt}

\begin{longtable}{lcccccccccc}
%{\centering\arraybackslash}p{0.08\textwidth}}
    %\caption{\makebox[\textwidth][c]{Your Table Caption Here}}
    %\caption{Extended Table \ref{tab:top_rank} for all the massive galaxies in our sample (with a SMBH mass $M_{\rm{BH}} >10^{8.4}M_\odot$, or equivalently with a chirp mass $M_{chirp}>10^8 M_\odot$). Columns include: galaxy name, total score rank, luminosity distance $D$, IFU survey, log of the black hole mass $M_{\rm{BH}}$, log of the hypothetical strain $h_0$, LDA score (Equation~\ref{eq:LDA_kin}), normalized log hypothetical GW strain $\widehat{\log h_0}$, normalized LDA score $\widehat{\rm{LDA}}$, total score (Equation~\ref{eq:score}), and the inner light profile classification from the literature (where `C'=core, `P'=power-law, and `I'=intermediate).} \label{tab:top_rank_full}
    \caption{Ranking of all the massive galaxies in our sample (with $M_{\rm{BH}} >10^{8.4}M_\odot$). Columns include: galaxy name, total score rank, luminosity distance $D$, IFU survey, log of the black hole mass $M_{\rm{BH}}$, log of the hypothetical GW strain $h_0$, LDA score (Equation~\ref{eq:LDA_kin}), normalized log of the hypothetical GW strain $\widehat{\log h_0}$, normalized LDA score $\widehat{\rm{LDA}}$, total score (Equation~\ref{eq:score}), and the inner light profile classification from the literature (where `C'=core, `P'=power-law, and `I'=intermediate).} \label{tab:top_rank_full}\\

    % Header Row
    \hline
    \hline
    \multicolumn{1}{c}{Name} &
    \multicolumn{1}{c}{Total Score} &
    \multicolumn{1}{c}{$D$} &
    \multicolumn{1}{c}{Survey} &
    \multicolumn{1}{c}{$\log M_{\rm{BH}}$} &
    \multicolumn{1}{c}{$\log h_0$} &
    \multicolumn{1}{c}{LDA} &
    \multicolumn{1}{c}{$\widehat{\log h_0}$} &
    \multicolumn{1}{c}{$\widehat{\rm{LDA}}$} &
    \multicolumn{1}{c}{Total} &
    \multicolumn{1}{c}{Light} \\
    \multicolumn{1}{c}{} &
    \multicolumn{1}{c}{Rank} &
    \multicolumn{1}{c}{[Mpc]} &
    \multicolumn{1}{c}{} &
    \multicolumn{1}{c}{$[M_\odot]$} &
    \multicolumn{1}{c}{} &
    \multicolumn{1}{c}{Score} &
    \multicolumn{1}{c}{} &
    \multicolumn{1}{c}{Score} &
    \multicolumn{1}{c}{Score} &
    \multicolumn{1}{c}{Profile} \\
    \hline
    \hline
    \endfirsthead

    % Repeating Header for Subsequent Pages
    \hline
    \hline
    \multicolumn{1}{c}{Name} &
    \multicolumn{1}{c}{Rank} &
    \multicolumn{1}{c}{$D$} &
    \multicolumn{1}{c}{Survey} &
    \multicolumn{1}{c}{$\log M_{\rm{BH}}$} &
    \multicolumn{1}{c}{$\log h_0$} &
    \multicolumn{1}{c}{LDA} &
    \multicolumn{1}{c}{$\widehat{\log h_0}$} &
    \multicolumn{1}{c}{$\widehat{\rm{LDA}}$} &
    \multicolumn{1}{c}{Total} &
    \multicolumn{1}{c}{Light} \\
    \multicolumn{1}{c}{} &
    \multicolumn{1}{c}{} &
    \multicolumn{1}{c}{[Mpc]} &
    \multicolumn{1}{c}{} &
    \multicolumn{1}{c}{$[M_\odot]$} &
    \multicolumn{1}{c}{} &
    \multicolumn{1}{c}{} &
    \multicolumn{1}{c}{} &
    \multicolumn{1}{c}{} &
    \multicolumn{1}{c}{Score} &
    \multicolumn{1}{c}{Profile} \\
    \hline
    \hline
    \endhead

    % Footer on every page except the last
    \hline
    \multicolumn{11}{r}{\textit{Continued on next page}} \\
    \hline
    \endfoot

    % Footer on the last page
    \hline
    \endlastfoot

    % Example rows
    NGC4073 & 1 & 91.50 & MASSIVE & 10.42 & -12.76 & 4.84 & 0.97 & 0.99 & 0.98 & C \\
    NGC4486 & 2 & 17.20 & ATLAS3D & 9.97 & -12.78 & 4.80 & 0.96 & 0.98 & 0.97 & C \\
    NGC1016 & 3 & 95.20 & MASSIVE & 10.42 & -12.77 & 4.62 & 0.96 & 0.97 & 0.97 & C \\
    NGC2832 & 4 & 105.20 & MASSIVE & 10.46 & -12.75 & 4.41 & 0.97 & 0.95 & 0.96 & C \\
    NGC1060 & 5 & 67.40 & MASSIVE & 10.21 & -12.97 & 4.97 & 0.91 & 1.00 & 0.95 & -- \\
    NGC4472 & 6 & 17.10 & ATLAS3D & 10.04 & -12.67 & 3.80 & 0.99 & 0.89 & 0.94 & C \\
    NGC0533 & 7 & 77.90 & MASSIVE & 10.24 & -12.99 & 4.64 & 0.90 & 0.97 & 0.94 & -- \\
    NGC4874 & 8 & 102.00 & MASSIVE & 10.32 & -12.97 & 4.35 & 0.91 & 0.94 & 0.93 & C \\
    NGC0410 & 9 & 71.30 & MASSIVE & 10.15 & -13.09 & 4.70 & 0.87 & 0.97 & 0.92 & -- \\
    NGC4406 & 10 & 16.80 & ATLAS3D & 9.79 & -13.07 & 4.62 & 0.88 & 0.97 & 0.92 & C \\
    NGC7265 & 11 & 82.80 & MASSIVE & 10.17 & -13.13 & 4.78 & 0.86 & 0.98 & 0.92 & -- \\
    NGC4374 & 12 & 18.50 & ATLAS3D & 9.77 & -13.15 & 4.74 & 0.86 & 0.98 & 0.92 & C \\
    NGC1060 & 13 & 73.90 & CALIFA & 10.13 & -13.14 & 4.70 & 0.86 & 0.97 & 0.92 & -- \\
    NGC0777 & 14 & 72.20 & MASSIVE & 10.17 & -13.07 & 4.47 & 0.88 & 0.95 & 0.92 & -- \\
    NGC1129 & 15 & 73.90 & MASSIVE & 10.29 & -12.87 & 3.68 & 0.93 & 0.88 & 0.91 & -- \\
    NGC4261 & 16 & 30.80 & ATLAS3D & 9.96 & -13.05 & 4.21 & 0.88 & 0.93 & 0.91 & C \\
    NGC0507 & 17 & 69.80 & MASSIVE & 10.17 & -13.06 & 4.20 & 0.88 & 0.93 & 0.91 & C \\
    NGC5846 & 18 & 24.20 & ATLAS3D & 9.75 & -13.29 & 4.84 & 0.81 & 0.99 & 0.90 & C \\
    NGC2258 & 19 & 59.00 & MASSIVE & 10.00 & -13.27 & 4.73 & 0.82 & 0.98 & 0.90 & -- \\
    NGC0315 & 20 & 70.30 & MASSIVE & 10.39 & -12.69 & 2.88 & 0.99 & 0.81 & 0.90 & -- \\
    NGC0708 & 21 & 69.00 & MASSIVE & 10.00 & -13.33 & 4.80 & 0.80 & 0.98 & 0.90 & -- \\
    NGC4472 & 22 & 16.70 & MASSIVE & 10.04 & -12.65 & 2.43 & 1.00 & 0.77 & 0.89 & C \\
    NGC5557 & 23 & 51.00 & MASSIVE & 9.87 & -13.41 & 4.77 & 0.78 & 0.98 & 0.89 & C \\
    NGC2783 & 24 & 101.40 & MASSIVE & 10.04 & -13.43 & 4.71 & 0.77 & 0.98 & 0.88 & -- \\
    NGC7626 & 25 & 54.00 & MASSIVE & 10.00 & -13.23 & 4.11 & 0.83 & 0.92 & 0.88 & I \\
    NGC4636 & 26 & 14.30 & ATLAS3D & 9.50 & -13.48 & 4.69 & 0.76 & 0.97 & 0.87 & C \\
    NGC7436 & 27 & 106.60 & MASSIVE & 10.31 & -13.01 & 3.35 & 0.90 & 0.85 & 0.87 & -- \\
    NGC2274 & 28 & 73.80 & MASSIVE & 10.01 & -13.34 & 4.21 & 0.80 & 0.93 & 0.87 & -- \\
    NGC4914 & 29 & 74.50 & MASSIVE & 10.04 & -13.30 & 4.08 & 0.81 & 0.92 & 0.87 & -- \\
    NGC1573 & 30 & 65.00 & MASSIVE & 9.93 & -13.42 & 4.40 & 0.78 & 0.95 & 0.87 & -- \\
    NGC0080 & 31 & 81.90 & MASSIVE & 10.00 & -13.41 & 4.32 & 0.78 & 0.94 & 0.86 & -- \\
    NGC5322 & 32 & 34.20 & MASSIVE & 9.90 & -13.19 & 3.71 & 0.84 & 0.88 & 0.86 & C \\
    NGC3842 & 33 & 99.40 & MASSIVE & 10.15 & -13.24 & 3.81 & 0.83 & 0.89 & 0.86 & C \\
    NGC4365 & 34 & 23.30 & ATLAS3D & 9.68 & -13.39 & 4.19 & 0.79 & 0.93 & 0.86 & C \\
    NGC3209 & 35 & 94.60 & MASSIVE & 9.93 & -13.59 & 4.68 & 0.73 & 0.97 & 0.86 & -- \\
    NGC5813 & 36 & 31.30 & ATLAS3D & 9.78 & -13.36 & 3.96 & 0.79 & 0.91 & 0.85 & C \\
    NGC2672 & 37 & 61.50 & MASSIVE & 9.96 & -13.35 & 3.92 & 0.80 & 0.90 & 0.85 & -- \\
    NGC3562 & 38 & 101.00 & MASSIVE & 10.00 & -13.50 & 4.24 & 0.75 & 0.93 & 0.85 & -- \\
    NGC1132 & 39 & 97.60 & MASSIVE & 10.03 & -13.44 & 4.08 & 0.77 & 0.92 & 0.85 & -- \\
    NGC7436B & 40 & 107.80 & CALIFA & 10.23 & -13.14 & 3.22 & 0.86 & 0.84 & 0.85 & -- \\
    NGC4874 & 41 & 114.00 & CALIFA & 9.92 & -13.68 & 4.64 & 0.70 & 0.97 & 0.85 & C \\
    NGC4649 & 42 & 17.30 & ATLAS3D & 9.96 & -12.80 & 1.93 & 0.95 & 0.72 & 0.84 & C \\
    NGC0890 & 43 & 55.60 & MASSIVE & 9.90 & -13.40 & 3.88 & 0.78 & 0.90 & 0.84 & -- \\
    NGC0499 & 44 & 69.80 & MASSIVE & 9.90 & -13.50 & 4.09 & 0.75 & 0.92 & 0.84 & -- \\
    NGC3462 & 45 & 99.20 & MASSIVE & 9.97 & -13.54 & 4.12 & 0.74 & 0.92 & 0.84 & -- \\
    UGC02783 & 46 & 85.80 & MASSIVE & 9.86 & -13.66 & 4.41 & 0.71 & 0.95 & 0.84 & -- \\
    NGC7386 & 47 & 99.10 & MASSIVE & 9.96 & -13.56 & 4.10 & 0.74 & 0.92 & 0.83 & -- \\
    NGC0910 & 48 & 79.80 & MASSIVE & 9.80 & -13.72 & 4.45 & 0.69 & 0.95 & 0.83 & C \\
    NGC6173 & 49 & 136.70 & CALIFA & 9.97 & -13.69 & 4.26 & 0.70 & 0.93 & 0.83 & -- \\
    NGC7274 & 50 & 82.80 & MASSIVE & 9.83 & -13.69 & 4.20 & 0.70 & 0.93 & 0.82 & -- \\
    NGC6338 & 51 & 126.40 & CALIFA & 9.92 & -13.73 & 4.18 & 0.69 & 0.93 & 0.82 & -- \\
    NGC3937 & 52 & 101.20 & MASSIVE & 9.97 & -13.55 & 3.63 & 0.74 & 0.88 & 0.81 & -- \\
    NGC4555 & 53 & 103.60 & MASSIVE & 10.15 & -13.25 & 2.68 & 0.83 & 0.79 & 0.81 & -- \\
    NGC3158 & 54 & 103.40 & MASSIVE & 10.38 & -12.88 & 1.26 & 0.93 & 0.66 & 0.81 & -- \\
    NGC1684 & 55 & 63.50 & MASSIVE & 9.80 & -13.62 & 3.68 & 0.72 & 0.88 & 0.80 & -- \\
    NGC7556 & 56 & 103.00 & MASSIVE & 10.11 & -13.32 & 2.82 & 0.81 & 0.80 & 0.80 & -- \\
    NGC5557 & 57 & 38.80 & ATLAS3D & 9.42 & -14.06 & 4.62 & 0.59 & 0.97 & 0.80 & C \\
    IC0310 & 58 & 77.50 & MASSIVE & 9.80 & -13.71 & 3.82 & 0.69 & 0.89 & 0.80 & -- \\
    NGC4816 & 59 & 102.00 & MASSIVE & 9.80 & -13.83 & 4.08 & 0.66 & 0.92 & 0.80 & -- \\
    NGC7619 & 60 & 54.00 & MASSIVE & 10.00 & -13.23 & 2.36 & 0.83 & 0.76 & 0.80 & C \\
    NGC5322 & 61 & 30.30 & ATLAS3D & 9.69 & -13.48 & 3.09 & 0.76 & 0.83 & 0.79 & C \\
    UGC03683 & 62 & 85.10 & MASSIVE & 9.92 & -13.56 & 3.30 & 0.74 & 0.85 & 0.79 & -- \\
    NGC7550 & 63 & 73.80 & CALIFA & 9.55 & -14.10 & 4.49 & 0.58 & 0.96 & 0.79 & -- \\
    NGC7052 & 64 & 69.30 & MASSIVE & 10.00 & -13.34 & 2.57 & 0.80 & 0.78 & 0.79 & C \\
    NGC0741 & 65 & 78.70 & CALIFA & 9.67 & -13.93 & 4.03 & 0.63 & 0.91 & 0.78 & -- \\
    NGC4168 & 66 & 30.90 & ATLAS3D & 9.37 & -14.03 & 4.23 & 0.60 & 0.93 & 0.78 & C \\
    NGC4552 & 67 & 15.80 & ATLAS3D & 9.23 & -13.97 & 4.08 & 0.62 & 0.92 & 0.78 & C \\
    NGC2320 & 68 & 89.40 & MASSIVE & 10.17 & -13.17 & 1.79 & 0.85 & 0.71 & 0.78 & -- \\
    NGC2513 & 69 & 70.80 & MASSIVE & 9.92 & -13.48 & 2.84 & 0.76 & 0.80 & 0.78 & -- \\
    NGC6482 & 70 & 61.40 & MASSIVE & 9.96 & -13.35 & 2.43 & 0.80 & 0.77 & 0.78 & -- \\
    NGC4841A & 71 & 108.00 & CALIFA & 9.71 & -14.00 & 4.02 & 0.61 & 0.91 & 0.78 & -- \\
    NGC4382 & 72 & 17.90 & ATLAS3D & 9.58 & -13.45 & 2.62 & 0.77 & 0.78 & 0.77 & C \\
    NGC2513 & 73 & 71.10 & CALIFA & 9.70 & -13.84 & 3.56 & 0.66 & 0.87 & 0.77 & -- \\
    NGC1700 & 74 & 54.40 & MASSIVE & 9.96 & -13.30 & 1.86 & 0.81 & 0.71 & 0.76 & C \\
    UGC03894 & 75 & 97.20 & MASSIVE & 9.96 & -13.55 & 2.66 & 0.74 & 0.79 & 0.76 & -- \\
    UGC12127 & 76 & 121.70 & CALIFA & 9.47 & -14.47 & 4.62 & 0.47 & 0.97 & 0.76 & -- \\
    NGC3816 & 77 & 99.40 & MASSIVE & 9.85 & -13.75 & 3.16 & 0.68 & 0.83 & 0.76 & -- \\
    NGC4816 & 78 & 110.10 & CALIFA & 9.66 & -14.11 & 3.92 & 0.58 & 0.90 & 0.76 & -- \\
    NGC0810 & 79 & 109.40 & CALIFA & 9.72 & -14.00 & 3.59 & 0.61 & 0.87 & 0.75 & -- \\
    NGC6703 & 80 & 25.90 & ATLAS3D & 8.92 & -14.71 & 4.68 & 0.40 & 0.97 & 0.75 & P \\
    NGC5198 & 81 & 39.60 & ATLAS3D & 9.21 & -14.41 & 4.23 & 0.49 & 0.93 & 0.74 & C \\
    NGC0499 & 82 & 62.90 & CALIFA & 9.51 & -14.11 & 3.65 & 0.58 & 0.88 & 0.74 & -- \\
    NGC5490 & 83 & 78.60 & MASSIVE & 9.94 & -13.48 & 2.02 & 0.76 & 0.73 & 0.74 & -- \\
    NGC0383 & 84 & 71.30 & MASSIVE & 10.10 & -13.18 & 0.89 & 0.84 & 0.62 & 0.74 & -- \\
    NGC5029 & 85 & 136.00 & CALIFA & 9.65 & -14.21 & 3.82 & 0.55 & 0.89 & 0.74 & -- \\
    UGC10695 & 86 & 129.30 & CALIFA & 9.37 & -14.66 & 4.42 & 0.42 & 0.95 & 0.73 & -- \\
    NGC1453 & 87 & 56.40 & MASSIVE & 10.00 & -13.25 & 0.88 & 0.83 & 0.62 & 0.73 & -- \\
    NGC2693 & 88 & 74.40 & MASSIVE & 10.06 & -13.27 & 0.79 & 0.82 & 0.61 & 0.72 & -- \\
    NGC3158 & 89 & 107.90 & CALIFA & 9.98 & -13.56 & 1.81 & 0.74 & 0.71 & 0.72 & -- \\
    NGC6125 & 90 & 77.00 & CALIFA & 9.49 & -14.23 & 3.50 & 0.54 & 0.86 & 0.72 & -- \\
    NGC2768 & 91 & 21.80 & ATLAS3D & 9.70 & -13.34 & 0.96 & 0.80 & 0.63 & 0.72 & I \\
    IC1079 & 92 & 137.90 & CALIFA & 9.41 & -14.62 & 4.12 & 0.43 & 0.92 & 0.72 & -- \\
    NGC2418 & 93 & 74.10 & MASSIVE & 9.85 & -13.62 & 1.92 & 0.72 & 0.72 & 0.72 & -- \\
    NGC6575 & 94 & 106.00 & MASSIVE & 9.96 & -13.59 & 1.76 & 0.73 & 0.70 & 0.72 & -- \\
    NGC0997 & 95 & 90.40 & MASSIVE & 9.85 & -13.71 & 2.08 & 0.69 & 0.73 & 0.71 & -- \\
    UGC10693 & 96 & 129.80 & CALIFA & 9.66 & -14.17 & 3.21 & 0.56 & 0.84 & 0.71 & -- \\
    NGC6223 & 97 & 86.70 & MASSIVE & 9.96 & -13.50 & 1.24 & 0.75 & 0.65 & 0.71 & -- \\
    NGC6515 & 98 & 106.20 & CALIFA & 9.22 & -14.82 & 4.15 & 0.37 & 0.92 & 0.70 & -- \\
    NGC7562 & 99 & 52.70 & CALIFA & 9.30 & -14.39 & 3.45 & 0.50 & 0.86 & 0.70 & -- \\
    NGC6375 & 100 & 95.80 & MASSIVE & 9.92 & -13.62 & 1.47 & 0.72 & 0.68 & 0.70 & -- \\
    NGC6411 & 101 & 61.30 & CALIFA & 9.06 & -14.84 & 4.03 & 0.37 & 0.91 & 0.70 & -- \\
    NGC3608 & 102 & 22.30 & ATLAS3D & 8.89 & -14.70 & 3.76 & 0.41 & 0.89 & 0.69 & C \\
    NGC3607 & 103 & 22.20 & ATLAS3D & 9.43 & -13.79 & 1.86 & 0.67 & 0.71 & 0.69 & C \\
    NGC7618 & 104 & 76.30 & MASSIVE & 9.86 & -13.61 & 1.27 & 0.72 & 0.66 & 0.69 & -- \\
    NGC3414 & 105 & 24.50 & ATLAS3D & 9.10 & -14.39 & 3.23 & 0.50 & 0.84 & 0.69 & P \\
    NGC5831 & 106 & 26.40 & ATLAS3D & 8.77 & -14.97 & 4.07 & 0.33 & 0.92 & 0.69 & P \\
    NGC5614 & 107 & 65.40 & CALIFA & 9.37 & -14.36 & 3.16 & 0.50 & 0.83 & 0.69 & -- \\
    NGC6020 & 108 & 71.80 & CALIFA & 8.95 & -15.10 & 4.14 & 0.29 & 0.92 & 0.68 & -- \\
    NGC4278 & 109 & 15.60 & ATLAS3D & 9.06 & -14.26 & 2.88 & 0.53 & 0.81 & 0.68 & C \\
    NGC5485 & 110 & 25.20 & ATLAS3D & 9.03 & -14.52 & 3.32 & 0.46 & 0.85 & 0.68 & C \\
    NGC5129 & 111 & 107.50 & MASSIVE & 10.15 & -13.27 & -0.41 & 0.82 & 0.50 & 0.68 & -- \\
    NGC6146 & 112 & 137.30 & CALIFA & 9.83 & -13.91 & 1.67 & 0.63 & 0.69 & 0.67 & -- \\
    NGC5631 & 113 & 36.30 & CALIFA & 8.85 & -14.97 & 3.66 & 0.33 & 0.88 & 0.66 & -- \\
    NGC0155 & 114 & 89.20 & CALIFA & 9.20 & -14.78 & 3.35 & 0.38 & 0.85 & 0.66 & -- \\
    NGC0524 & 115 & 23.30 & ATLAS3D & 9.51 & -13.68 & 0.76 & 0.70 & 0.61 & 0.66 & C \\
    UGC00029 & 116 & 127.50 & CALIFA & 9.00 & -15.27 & 3.86 & 0.24 & 0.90 & 0.66 & -- \\
    NGC5631 & 117 & 27.00 & ATLAS3D & 8.79 & -14.94 & 3.51 & 0.34 & 0.86 & 0.66 & -- \\
    NGC1167 & 118 & 70.20 & MASSIVE & 9.99 & -13.36 & -0.69 & 0.79 & 0.48 & 0.65 & -- \\
    NGC3665 & 119 & 33.10 & ATLAS3D & 9.73 & -13.47 & -0.28 & 0.76 & 0.51 & 0.65 & -- \\
    NGC5485 & 120 & 36.70 & CALIFA & 8.98 & -14.75 & 3.12 & 0.39 & 0.83 & 0.65 & C \\
    NGC5576 & 121 & 24.80 & ATLAS3D & 8.78 & -14.92 & 3.32 & 0.34 & 0.85 & 0.65 & C \\
    NGC5481 & 122 & 25.80 & ATLAS3D & 8.41 & -15.56 & 3.88 & 0.16 & 0.90 & 0.65 & -- \\
    NGC3379 & 123 & 10.30 & ATLAS3D & 8.83 & -14.46 & 2.40 & 0.48 & 0.76 & 0.64 & C \\
    NGC3615 & 124 & 101.20 & MASSIVE & 9.96 & -13.57 & -0.36 & 0.73 & 0.51 & 0.63 & -- \\
    NGC3193 & 125 & 33.10 & ATLAS3D & 9.16 & -14.41 & 2.14 & 0.49 & 0.74 & 0.63 & C \\
    NGC0665 & 126 & 74.60 & MASSIVE & 9.90 & -13.53 & -0.65 & 0.75 & 0.48 & 0.63 & -- \\
    NGC3303 & 127 & 98.80 & CALIFA & 9.04 & -15.10 & 3.18 & 0.29 & 0.83 & 0.63 & -- \\
    NGC5353 & 128 & 41.10 & MASSIVE & 9.87 & -13.32 & -1.96 & 0.81 & 0.36 & 0.62 & -- \\
    NGC1167 & 129 & 70.60 & CALIFA & 9.92 & -13.48 & -1.00 & 0.76 & 0.45 & 0.62 & -- \\
    NGC4494 & 130 & 16.60 & ATLAS3D & 8.94 & -14.48 & 2.12 & 0.47 & 0.74 & 0.62 & P \\
    NGC3613 & 131 & 28.30 & ATLAS3D & 9.27 & -14.16 & 1.34 & 0.56 & 0.66 & 0.62 & C \\
    NGC7426 & 132 & 80.00 & MASSIVE & 10.06 & -13.30 & -2.53 & 0.81 & 0.31 & 0.61 & -- \\
    NGC3805 & 133 & 99.40 & MASSIVE & 10.01 & -13.47 & -1.42 & 0.76 & 0.41 & 0.61 & -- \\
    NGC7025 & 134 & 75.40 & CALIFA & 9.69 & -13.89 & 0.36 & 0.64 & 0.57 & 0.61 & -- \\
    NGC0474 & 135 & 30.90 & ATLAS3D & 8.85 & -14.89 & 2.50 & 0.35 & 0.77 & 0.60 & P \\
    NGC0661 & 136 & 30.60 & ATLAS3D & 8.85 & -14.89 & 2.47 & 0.35 & 0.77 & 0.60 & -- \\
    NGC7619 & 137 & 54.90 & CALIFA & 8.87 & -15.11 & 2.75 & 0.29 & 0.79 & 0.60 & C \\
    NGC2918 & 138 & 105.10 & CALIFA & 9.57 & -14.23 & 1.07 & 0.54 & 0.64 & 0.59 & -- \\
    NGC4621 & 139 & 14.90 & ATLAS3D & 9.12 & -14.14 & 0.77 & 0.57 & 0.61 & 0.59 & P \\
    NGC7454 & 140 & 23.20 & ATLAS3D & 8.43 & -15.48 & 2.88 & 0.18 & 0.81 & 0.58 & -- \\
    NGC3106 & 141 & 96.40 & CALIFA & 9.25 & -14.73 & 1.98 & 0.40 & 0.72 & 0.58 & -- \\
    UGC10097 & 142 & 94.60 & CALIFA & 9.59 & -14.15 & 0.65 & 0.57 & 0.60 & 0.58 & -- \\
    NGC4753 & 143 & 22.90 & ATLAS3D & 9.49 & -13.70 & -1.10 & 0.70 & 0.44 & 0.58 & -- \\
    NGC7623 & 144 & 54.50 & CALIFA & 8.92 & -15.03 & 2.34 & 0.31 & 0.76 & 0.58 & -- \\
    NGC3226 & 145 & 22.90 & ATLAS3D & 8.94 & -14.62 & 1.69 & 0.43 & 0.70 & 0.58 & P \\
    NGC4477 & 146 & 16.50 & ATLAS3D & 8.87 & -14.59 & 1.52 & 0.44 & 0.68 & 0.57 & I \\
    NGC1497 & 147 & 87.80 & MASSIVE & 9.79 & -13.79 & -1.00 & 0.67 & 0.45 & 0.57 & -- \\
    NGC4191 & 148 & 39.20 & ATLAS3D & 8.54 & -15.53 & 2.68 & 0.17 & 0.79 & 0.57 & -- \\
    NGC4643 & 149 & 16.50 & ATLAS3D & 8.80 & -14.71 & 1.64 & 0.40 & 0.69 & 0.57 & -- \\
    NGC4429 & 150 & 16.50 & ATLAS3D & 9.18 & -14.07 & -0.05 & 0.59 & 0.54 & 0.56 & P \\
    NGC4526 & 151 & 16.40 & ATLAS3D & 9.29 & -13.89 & -0.78 & 0.64 & 0.47 & 0.56 & -- \\
    NGC0821 & 152 & 23.40 & ATLAS3D & 9.08 & -14.39 & 0.84 & 0.50 & 0.62 & 0.56 & I \\
    NGC5638 & 153 & 25.60 & ATLAS3D & 8.85 & -14.82 & 1.64 & 0.37 & 0.69 & 0.56 & -- \\
    NGC3640 & 154 & 26.30 & ATLAS3D & 9.26 & -14.14 & -0.08 & 0.57 & 0.53 & 0.55 & C \\
    NGC0447 & 155 & 80.10 & CALIFA & 9.13 & -14.84 & 1.57 & 0.37 & 0.69 & 0.55 & -- \\
    NGC1289 & 156 & 38.40 & ATLAS3D & 8.55 & -15.49 & 2.31 & 0.18 & 0.75 & 0.55 & -- \\
    NGC4697 & 157 & 11.40 & ATLAS3D & 9.05 & -14.14 & -0.16 & 0.57 & 0.53 & 0.55 & P \\
    NGC7194 & 158 & 118.90 & CALIFA & 9.57 & -14.28 & 0.23 & 0.53 & 0.56 & 0.54 & -- \\
    NGC3615 & 159 & 105.40 & CALIFA & 9.49 & -14.37 & 0.44 & 0.50 & 0.58 & 0.54 & -- \\
    NGC5353 & 160 & 35.20 & ATLAS3D & 9.65 & -13.62 & -3.01 & 0.72 & 0.26 & 0.54 & -- \\
    NGC5966 & 161 & 73.50 & CALIFA & 8.96 & -15.09 & 1.76 & 0.29 & 0.70 & 0.54 & -- \\
    NGC5208 & 162 & 105.00 & MASSIVE & 9.97 & -13.56 & -3.82 & 0.74 & 0.19 & 0.54 & -- \\
    NGC4596 & 163 & 16.50 & ATLAS3D & 8.83 & -14.66 & 0.99 & 0.42 & 0.63 & 0.54 & P \\
    NGC0169 & 164 & 66.20 & CALIFA & 9.79 & -13.66 & -2.89 & 0.71 & 0.27 & 0.54 & -- \\
    NGC4608 & 165 & 16.50 & ATLAS3D & 8.42 & -15.35 & 1.98 & 0.22 & 0.72 & 0.53 & -- \\
    NGC5866 & 166 & 14.90 & ATLAS3D & 8.95 & -14.42 & 0.37 & 0.49 & 0.57 & 0.53 & -- \\
    NGC4473 & 167 & 15.30 & ATLAS3D & 8.85 & -14.60 & 0.77 & 0.44 & 0.61 & 0.53 & C \\
    NGC2554 & 168 & 64.00 & CALIFA & 9.25 & -14.55 & 0.61 & 0.45 & 0.60 & 0.53 & -- \\
    NGC7563 & 169 & 60.80 & CALIFA & 8.90 & -15.12 & 1.57 & 0.29 & 0.69 & 0.53 & -- \\
    NGC4690 & 170 & 40.20 & ATLAS3D & 8.42 & -15.74 & 2.04 & 0.11 & 0.73 & 0.52 & -- \\
    NGC0932 & 171 & 57.50 & CALIFA & 8.90 & -15.09 & 1.37 & 0.30 & 0.67 & 0.52 & -- \\
    NGC6945 & 172 & 59.30 & CALIFA & 9.49 & -14.11 & -1.11 & 0.58 & 0.44 & 0.51 & -- \\
    NGC5784 & 173 & 86.60 & CALIFA & 9.25 & -14.68 & 0.59 & 0.41 & 0.59 & 0.51 & -- \\
    NGC3619 & 174 & 26.80 & ATLAS3D & 8.82 & -14.89 & 0.95 & 0.35 & 0.63 & 0.51 & -- \\
    NGC0529 & 175 & 69.00 & CALIFA & 9.07 & -14.88 & 0.79 & 0.36 & 0.61 & 0.50 & -- \\
    NGC0936 & 176 & 22.40 & ATLAS3D & 9.39 & -13.86 & -2.99 & 0.65 & 0.26 & 0.49 & P \\
    NGC7722 & 177 & 58.30 & CALIFA & 9.29 & -14.44 & -0.36 & 0.48 & 0.51 & 0.49 & -- \\
    NGC6314 & 178 & 105.30 & CALIFA & 9.25 & -14.77 & 0.37 & 0.39 & 0.57 & 0.49 & -- \\
    IC0719 & 179 & 29.40 & ATLAS3D & 8.44 & -15.56 & 1.42 & 0.16 & 0.67 & 0.49 & -- \\
    UGC10905 & 180 & 122.70 & CALIFA & 9.80 & -13.92 & -2.92 & 0.63 & 0.27 & 0.49 & -- \\
    NGC5876 & 181 & 55.00 & CALIFA & 8.81 & -15.22 & 1.05 & 0.26 & 0.64 & 0.49 & -- \\
    NGC0680 & 182 & 37.50 & ATLAS3D & 8.99 & -14.76 & 0.20 & 0.39 & 0.56 & 0.48 & -- \\
    NGC2859 & 183 & 27.00 & ATLAS3D & 8.91 & -14.74 & 0.14 & 0.40 & 0.55 & 0.48 & P \\
    NGC5947 & 184 & 86.30 & CALIFA & 8.75 & -15.52 & 1.23 & 0.17 & 0.65 & 0.48 & -- \\
    NGC4624 & 185 & 16.50 & ATLAS3D & 8.77 & -14.77 & 0.13 & 0.39 & 0.55 & 0.48 & -- \\
    NGC6021 & 186 & 78.50 & CALIFA & 8.96 & -15.13 & 0.74 & 0.28 & 0.61 & 0.48 & -- \\
    NGC1023 & 187 & 11.10 & ATLAS3D & 8.70 & -14.71 & -0.07 & 0.40 & 0.53 & 0.47 & P \\
    NGC4442 & 188 & 15.30 & ATLAS3D & 8.69 & -14.86 & 0.23 & 0.36 & 0.56 & 0.47 & P \\
    NGC6548 & 189 & 22.40 & ATLAS3D & 8.77 & -14.90 & 0.24 & 0.35 & 0.56 & 0.47 & -- \\
    NGC2962 & 190 & 34.00 & ATLAS3D & 9.09 & -14.55 & -0.82 & 0.45 & 0.46 & 0.46 & P \\
    NGC7738 & 191 & 97.80 & CALIFA & 9.06 & -15.05 & 0.28 & 0.31 & 0.57 & 0.45 & -- \\
    NGC5473 & 192 & 33.20 & ATLAS3D & 9.07 & -14.56 & -0.85 & 0.45 & 0.46 & 0.45 & -- \\
    NGC4459 & 193 & 16.10 & ATLAS3D & 8.84 & -14.64 & -0.63 & 0.42 & 0.48 & 0.45 & P \\
    NGC5838 & 194 & 21.80 & ATLAS3D & 9.17 & -14.22 & -2.21 & 0.55 & 0.34 & 0.45 & P \\
    NGC5000 & 195 & 90.80 & CALIFA & 8.57 & -15.83 & 1.02 & 0.08 & 0.63 & 0.45 & -- \\
    NGC3182 & 196 & 34.00 & ATLAS3D & 8.51 & -15.50 & 0.79 & 0.18 & 0.61 & 0.45 & -- \\
    NGC3998 & 197 & 13.70 & ATLAS3D & 8.86 & -14.53 & -1.11 & 0.46 & 0.44 & 0.45 & I \\
    NGC7824 & 198 & 88.10 & CALIFA & 9.29 & -14.62 & -0.91 & 0.43 & 0.46 & 0.44 & -- \\
    UGC11228 & 199 & 90.30 & CALIFA & 9.08 & -14.98 & -0.04 & 0.33 & 0.54 & 0.44 & -- \\
    NGC4281 & 200 & 24.40 & ATLAS3D & 9.26 & -14.12 & -3.24 & 0.57 & 0.24 & 0.44 & P \\
    UGC05771 & 201 & 114.30 & CALIFA & 9.39 & -14.56 & -1.19 & 0.45 & 0.43 & 0.44 & -- \\
    NGC6081 & 202 & 85.00 & CALIFA & 9.12 & -14.90 & -0.31 & 0.35 & 0.51 & 0.44 & -- \\
    NGC1349 & 203 & 93.90 & CALIFA & 8.85 & -15.39 & 0.44 & 0.21 & 0.58 & 0.44 & -- \\
    NGC4956 & 204 & 77.70 & CALIFA & 8.93 & -15.17 & 0.15 & 0.27 & 0.55 & 0.44 & -- \\
    NGC6150 & 205 & 135.80 & CALIFA & 9.55 & -14.38 & -2.11 & 0.50 & 0.34 & 0.43 & -- \\
    NGC0023 & 206 & 65.70 & CALIFA & 9.01 & -14.97 & -0.46 & 0.33 & 0.50 & 0.42 & -- \\
    NGC7711 & 207 & 58.80 & CALIFA & 9.02 & -14.89 & -0.69 & 0.35 & 0.48 & 0.42 & -- \\
    NGC0776 & 208 & 69.70 & CALIFA & 8.52 & -15.80 & 0.44 & 0.09 & 0.58 & 0.42 & -- \\
    NGC5869 & 209 & 24.90 & ATLAS3D & 8.78 & -14.92 & -0.74 & 0.35 & 0.47 & 0.41 & -- \\
    UGC08107 & 210 & 128.60 & CALIFA & 9.04 & -15.20 & -0.22 & 0.26 & 0.52 & 0.41 & -- \\
    NGC2974 & 211 & 20.90 & ATLAS3D & 9.14 & -14.25 & -3.61 & 0.54 & 0.21 & 0.41 & P \\
    NGC4754 & 212 & 16.10 & ATLAS3D & 8.69 & -14.89 & -0.96 & 0.35 & 0.45 & 0.40 & P \\
    NGC6278 & 213 & 48.60 & CALIFA & 8.84 & -15.12 & -0.49 & 0.29 & 0.49 & 0.40 & P \\
    NGC5687 & 214 & 27.20 & ATLAS3D & 8.91 & -14.75 & -1.37 & 0.39 & 0.41 & 0.40 & -- \\
    NGC4270 & 215 & 35.20 & ATLAS3D & 8.59 & -15.39 & -0.14 & 0.21 & 0.53 & 0.40 & I \\
    UGC01271 & 216 & 71.90 & CALIFA & 8.71 & -15.51 & -0.03 & 0.17 & 0.54 & 0.40 & -- \\
    NGC5308 & 217 & 31.50 & ATLAS3D & 9.17 & -14.38 & -3.32 & 0.50 & 0.23 & 0.39 & P \\
    NGC5987 & 218 & 50.80 & CALIFA & 9.24 & -14.46 & -2.83 & 0.48 & 0.28 & 0.39 & -- \\
    NGC3945 & 219 & 23.20 & ATLAS3D & 8.98 & -14.56 & -2.37 & 0.45 & 0.32 & 0.39 & P \\
    NGC4036 & 220 & 24.60 & ATLAS3D & 9.18 & -14.26 & -4.41 & 0.53 & 0.13 & 0.39 & -- \\
    NGC0160 & 221 & 75.20 & CALIFA & 8.99 & -15.05 & -0.97 & 0.31 & 0.45 & 0.39 & -- \\
    NGC5218 & 222 & 49.30 & CALIFA & 8.46 & -15.75 & -0.16 & 0.10 & 0.52 & 0.38 & -- \\
    NGC4233 & 223 & 33.90 & ATLAS3D & 9.05 & -14.61 & -2.46 & 0.43 & 0.31 & 0.38 & -- \\
    NGC7671 & 224 & 60.00 & CALIFA & 8.89 & -15.12 & -0.96 & 0.28 & 0.45 & 0.38 & -- \\
    NGC2592 & 225 & 33.60 & CALIFA & 8.42 & -15.66 & -0.26 & 0.13 & 0.52 & 0.38 & P \\
    UGC10205 & 226 & 104.60 & CALIFA & 8.95 & -15.27 & -0.72 & 0.24 & 0.47 & 0.38 & -- \\
    NGC3230 & 227 & 40.80 & ATLAS3D & 9.11 & -14.59 & -2.69 & 0.44 & 0.29 & 0.37 & -- \\
    NGC5406 & 228 & 84.40 & CALIFA & 9.33 & -14.53 & -3.08 & 0.46 & 0.25 & 0.37 & -- \\
    NGC4371 & 229 & 17.00 & ATLAS3D & 8.69 & -14.92 & -1.61 & 0.35 & 0.39 & 0.37 & I \\
    NGC4762 & 230 & 22.60 & ATLAS3D & 9.10 & -14.35 & -4.64 & 0.51 & 0.11 & 0.37 & P \\
    NGC4268 & 231 & 31.70 & ATLAS3D & 8.59 & -15.35 & -0.78 & 0.22 & 0.47 & 0.37 & -- \\
    NGC6427 & 232 & 54.50 & CALIFA & 8.60 & -15.56 & -0.53 & 0.16 & 0.49 & 0.36 & -- \\
    UGC05113 & 233 & 103.60 & CALIFA & 9.09 & -15.02 & -1.46 & 0.31 & 0.40 & 0.36 & -- \\
    NGC7611 & 234 & 47.60 & CALIFA & 8.81 & -15.16 & -1.19 & 0.28 & 0.43 & 0.36 & -- \\
    UGC08234 & 235 & 124.80 & CALIFA & 9.14 & -15.03 & -1.52 & 0.31 & 0.40 & 0.36 & -- \\
    NGC5908 & 236 & 55.40 & CALIFA & 9.26 & -14.47 & -4.09 & 0.47 & 0.16 & 0.35 & -- \\
    NGC2639 & 237 & 52.70 & CALIFA & 9.19 & -14.57 & -3.37 & 0.44 & 0.23 & 0.35 & -- \\
    NGC4521 & 238 & 39.70 & ATLAS3D & 9.12 & -14.56 & -3.58 & 0.45 & 0.21 & 0.35 & -- \\
    UGC05108 & 239 & 123.70 & CALIFA & 8.79 & -15.60 & -0.77 & 0.15 & 0.47 & 0.35 & -- \\
    NGC2553 & 240 & 72.20 & CALIFA & 8.72 & -15.48 & -0.90 & 0.18 & 0.46 & 0.35 & -- \\
    NGC4676A & 241 & 105.30 & CALIFA & 8.69 & -15.70 & -0.70 & 0.12 & 0.48 & 0.35 & -- \\
    NGC0217 & 242 & 56.70 & CALIFA & 9.09 & -14.77 & -2.66 & 0.39 & 0.29 & 0.34 & -- \\
    UGC06036 & 243 & 101.60 & CALIFA & 9.19 & -14.85 & -2.39 & 0.36 & 0.32 & 0.34 & -- \\
    NGC4503 & 244 & 16.50 & ATLAS3D & 8.51 & -15.19 & -1.55 & 0.27 & 0.40 & 0.34 & P \\
    UGC06062 & 245 & 38.70 & ATLAS3D & 8.43 & -15.70 & -0.87 & 0.12 & 0.46 & 0.34 & P \\
    NGC4078 & 246 & 38.10 & ATLAS3D & 8.69 & -15.25 & -1.52 & 0.25 & 0.40 & 0.33 & -- \\
    NGC7683 & 247 & 54.20 & CALIFA & 8.98 & -14.93 & -2.34 & 0.34 & 0.32 & 0.33 & -- \\
    IC0944 & 248 & 112.40 & CALIFA & 9.31 & -14.69 & -3.44 & 0.41 & 0.22 & 0.33 & -- \\
    NGC0171 & 249 & 56.10 & CALIFA & 8.56 & -15.64 & -1.03 & 0.13 & 0.44 & 0.33 & -- \\
    NGC4003 & 250 & 103.60 & CALIFA & 9.05 & -15.09 & -1.98 & 0.30 & 0.36 & 0.33 & -- \\
    NGC5422 & 251 & 30.80 & ATLAS3D & 8.88 & -14.85 & -2.79 & 0.36 & 0.28 & 0.33 & I \\
    NGC2695 & 252 & 31.50 & ATLAS3D & 8.86 & -14.89 & -2.67 & 0.35 & 0.29 & 0.32 & -- \\
    UGC06312 & 253 & 100.70 & CALIFA & 8.99 & -15.18 & -1.89 & 0.27 & 0.37 & 0.32 & -- \\
    NGC6278 & 254 & 42.90 & ATLAS3D & 8.98 & -14.82 & -3.05 & 0.37 & 0.26 & 0.32 & P \\
    NGC4143 & 255 & 15.50 & ATLAS3D & 8.47 & -15.23 & -1.85 & 0.25 & 0.37 & 0.32 & P \\
    NGC3458 & 256 & 30.90 & ATLAS3D & 8.41 & -15.63 & -1.23 & 0.14 & 0.43 & 0.32 & P \\
    NGC0517 & 257 & 60.10 & CALIFA & 8.70 & -15.44 & -1.51 & 0.19 & 0.40 & 0.31 & -- \\
    NGC5797 & 258 & 65.60 & CALIFA & 8.73 & -15.42 & -1.54 & 0.20 & 0.40 & 0.31 & -- \\
    NGC6798 & 259 & 37.50 & ATLAS3D & 8.51 & -15.55 & -1.38 & 0.16 & 0.41 & 0.31 & -- \\
    NGC0001 & 260 & 65.60 & CALIFA & 8.67 & -15.53 & -1.42 & 0.17 & 0.41 & 0.31 & -- \\
    NGC2592 & 261 & 25.00 & ATLAS3D & 8.50 & -15.39 & -1.70 & 0.21 & 0.38 & 0.31 & P \\
    NGC5582 & 262 & 27.70 & ATLAS3D & 8.76 & -15.01 & -2.65 & 0.32 & 0.29 & 0.31 & -- \\
    NGC0774 & 263 & 65.20 & CALIFA & 8.84 & -15.24 & -2.03 & 0.25 & 0.35 & 0.31 & -- \\
    UGC00036 & 264 & 90.50 & CALIFA & 8.95 & -15.20 & -2.13 & 0.26 & 0.34 & 0.31 & -- \\
    NGC7311 & 265 & 66.90 & CALIFA & 9.05 & -14.91 & -3.19 & 0.35 & 0.24 & 0.30 & -- \\
    NGC6497 & 266 & 95.50 & CALIFA & 9.00 & -15.14 & -2.42 & 0.28 & 0.32 & 0.30 & -- \\
    NGC3595 & 267 & 34.70 & ATLAS3D & 8.48 & -15.56 & -1.60 & 0.16 & 0.39 & 0.30 & P \\
    UGC12274 & 268 & 112.10 & CALIFA & 9.16 & -14.94 & -3.15 & 0.34 & 0.25 & 0.30 & -- \\
    NGC3610 & 269 & 20.80 & ATLAS3D & 8.59 & -15.16 & -2.44 & 0.27 & 0.31 & 0.30 & P \\
    UGC08781 & 270 & 120.90 & CALIFA & 9.03 & -15.20 & -2.34 & 0.26 & 0.32 & 0.29 & -- \\
    UGC10380 & 271 & 137.90 & CALIFA & 8.96 & -15.36 & -1.99 & 0.22 & 0.36 & 0.29 & -- \\
    NGC1645 & 272 & 70.20 & CALIFA & 8.71 & -15.48 & -1.79 & 0.18 & 0.37 & 0.29 & -- \\
    NGC3674 & 273 & 33.40 & ATLAS3D & 8.71 & -15.17 & -2.44 & 0.27 & 0.31 & 0.29 & -- \\
    NGC2698 & 274 & 27.10 & ATLAS3D & 8.70 & -15.09 & -2.73 & 0.30 & 0.29 & 0.29 & -- \\
    NGC4179 & 275 & 16.50 & ATLAS3D & 8.60 & -15.04 & -2.91 & 0.31 & 0.27 & 0.29 & -- \\
    NGC3245 & 276 & 20.30 & ATLAS3D & 8.68 & -15.00 & -3.10 & 0.32 & 0.25 & 0.29 & P \\
    NGC4215 & 277 & 31.50 & ATLAS3D & 8.58 & -15.36 & -2.10 & 0.22 & 0.34 & 0.29 & -- \\
    NGC7684 & 278 & 74.30 & CALIFA & 8.93 & -15.15 & -2.67 & 0.28 & 0.29 & 0.29 & -- \\
    NGC4570 & 279 & 17.10 & ATLAS3D & 8.62 & -15.03 & -3.05 & 0.31 & 0.26 & 0.29 & P \\
    NGC5507 & 280 & 28.50 & ATLAS3D & 8.57 & -15.33 & -2.25 & 0.23 & 0.33 & 0.28 & -- \\
    NGC0364 & 281 & 72.80 & CALIFA & 8.90 & -15.20 & -2.64 & 0.26 & 0.30 & 0.28 & -- \\
    NGC4435 & 282 & 16.70 & ATLAS3D & 8.51 & -15.20 & -2.66 & 0.26 & 0.29 & 0.28 & -- \\
    NGC0429 & 283 & 80.00 & CALIFA & 8.66 & -15.63 & -1.85 & 0.14 & 0.37 & 0.28 & -- \\
    UGC09537 & 284 & 138.50 & CALIFA & 9.26 & -14.87 & -4.08 & 0.36 & 0.16 & 0.28 & -- \\
    NGC2577 & 285 & 30.80 & ATLAS3D & 8.84 & -14.92 & -3.85 & 0.34 & 0.18 & 0.28 & -- \\
    UGC03995 & 286 & 71.80 & CALIFA & 8.84 & -15.28 & -2.52 & 0.24 & 0.31 & 0.28 & -- \\
    NGC3300 & 287 & 51.00 & CALIFA & 8.62 & -15.51 & -2.08 & 0.17 & 0.35 & 0.27 & -- \\
    NGC5864 & 288 & 29.00 & ATLAS3D & 8.54 & -15.39 & -2.30 & 0.21 & 0.33 & 0.27 & -- \\
    NGC4546 & 289 & 13.70 & ATLAS3D & 8.61 & -14.95 & -3.76 & 0.34 & 0.19 & 0.27 & P \\
    NGC2880 & 290 & 28.70 & CALIFA & 8.49 & -15.47 & -2.17 & 0.19 & 0.34 & 0.27 & P \\
    UGC02229 & 291 & 104.30 & CALIFA & 8.80 & -15.52 & -2.10 & 0.17 & 0.35 & 0.27 & -- \\
    NGC7364 & 292 & 71.50 & CALIFA & 8.78 & -15.37 & -2.37 & 0.21 & 0.32 & 0.27 & -- \\
    NGC5888 & 293 & 136.50 & CALIFA & 9.24 & -14.90 & -4.11 & 0.35 & 0.16 & 0.27 & -- \\
    IC4566 & 294 & 92.30 & CALIFA & 8.89 & -15.32 & -2.56 & 0.23 & 0.30 & 0.27 & -- \\
    UGC10388 & 295 & 76.50 & CALIFA & 8.69 & -15.56 & -2.14 & 0.16 & 0.34 & 0.27 & -- \\
    IC0674 & 296 & 116.70 & CALIFA & 8.82 & -15.53 & -2.21 & 0.17 & 0.33 & 0.27 & -- \\
    NGC4350 & 297 & 15.40 & ATLAS3D & 8.56 & -15.09 & -3.35 & 0.30 & 0.23 & 0.26 & I \\
    NGC7787 & 298 & 96.10 & CALIFA & 8.42 & -16.11 & -1.80 & 0.00 & 0.37 & 0.26 & -- \\
    NGC5493 & 299 & 38.80 & ATLAS3D & 8.90 & -14.91 & -4.33 & 0.35 & 0.14 & 0.26 & -- \\
    NGC4251 & 300 & 19.10 & ATLAS3D & 8.41 & -15.42 & -2.65 & 0.20 & 0.29 & 0.25 & -- \\
    UGC11717 & 301 & 94.50 & CALIFA & 8.73 & -15.59 & -2.44 & 0.15 & 0.31 & 0.25 & -- \\
    NGC6547 & 302 & 40.80 & ATLAS3D & 8.83 & -15.05 & -4.02 & 0.31 & 0.17 & 0.25 & -- \\
    NGC4710 & 303 & 16.50 & ATLAS3D & 8.62 & -15.02 & -4.24 & 0.32 & 0.15 & 0.25 & -- \\
    NGC3160 & 304 & 106.90 & CALIFA & 8.89 & -15.38 & -2.88 & 0.21 & 0.27 & 0.24 & -- \\
    NGC3658 & 305 & 32.70 & ATLAS3D & 8.50 & -15.50 & -2.70 & 0.18 & 0.29 & 0.24 & -- \\
    NGC4111 & 306 & 14.60 & ATLAS3D & 8.42 & -15.28 & -3.24 & 0.24 & 0.24 & 0.24 & -- \\
    NGC6010 & 307 & 30.60 & ATLAS3D & 8.69 & -15.16 & -3.71 & 0.28 & 0.20 & 0.24 & -- \\
    NGC6004 & 308 & 64.80 & CALIFA & 8.51 & -15.79 & -2.33 & 0.09 & 0.32 & 0.24 & -- \\
    UGC02222 & 309 & 72.00 & CALIFA & 8.59 & -15.70 & -2.51 & 0.12 & 0.31 & 0.23 & -- \\
    NGC7591 & 310 & 72.00 & CALIFA & 8.61 & -15.66 & -2.59 & 0.13 & 0.30 & 0.23 & -- \\
    NGC2880 & 311 & 21.30 & ATLAS3D & 8.41 & -15.47 & -2.95 & 0.19 & 0.27 & 0.23 & P \\
    NGC0234 & 312 & 63.40 & CALIFA & 8.46 & -15.86 & -2.44 & 0.07 & 0.31 & 0.23 & -- \\
    NGC0036 & 313 & 86.60 & CALIFA & 8.80 & -15.42 & -3.11 & 0.20 & 0.25 & 0.23 & -- \\
    NGC4255 & 314 & 31.20 & ATLAS3D & 8.55 & -15.40 & -3.23 & 0.20 & 0.24 & 0.22 & -- \\
    NGC0528 & 315 & 68.70 & CALIFA & 8.77 & -15.38 & -3.33 & 0.21 & 0.23 & 0.22 & -- \\
    NGC6060 & 316 & 73.70 & CALIFA & 8.86 & -15.27 & -3.70 & 0.24 & 0.20 & 0.22 & -- \\
    IC1755 & 317 & 113.50 & CALIFA & 8.85 & -15.47 & -3.16 & 0.18 & 0.25 & 0.22 & -- \\
    IC1652 & 318 & 74.00 & CALIFA & 8.41 & -16.02 & -2.58 & 0.03 & 0.30 & 0.21 & -- \\
    NGC5934 & 319 & 91.30 & CALIFA & 8.87 & -15.34 & -3.65 & 0.22 & 0.20 & 0.21 & -- \\
    NGC3648 & 320 & 31.90 & ATLAS3D & 8.58 & -15.36 & -3.67 & 0.22 & 0.20 & 0.21 & -- \\
    NGC6941 & 321 & 94.80 & CALIFA & 8.87 & -15.35 & -3.68 & 0.22 & 0.20 & 0.21 & -- \\
    IC5376 & 322 & 72.90 & CALIFA & 8.47 & -15.91 & -2.75 & 0.06 & 0.29 & 0.21 & -- \\
    NGC6978 & 323 & 91.60 & CALIFA & 9.00 & -15.13 & -5.19 & 0.28 & 0.06 & 0.20 & -- \\
    NGC0192 & 324 & 58.90 & CALIFA & 8.74 & -15.37 & -3.83 & 0.21 & 0.19 & 0.20 & -- \\
    UGC10811 & 325 & 134.10 & CALIFA & 8.77 & -15.67 & -3.15 & 0.13 & 0.25 & 0.20 & -- \\
    NGC2347 & 326 & 67.10 & CALIFA & 8.87 & -15.21 & -4.73 & 0.26 & 0.10 & 0.20 & -- \\
    NGC0180 & 327 & 75.30 & CALIFA & 8.84 & -15.30 & -4.24 & 0.23 & 0.15 & 0.20 & -- \\
    UGC12185 & 328 & 97.50 & CALIFA & 8.49 & -16.00 & -2.88 & 0.03 & 0.27 & 0.19 & -- \\
    NGC7321 & 329 & 104.90 & CALIFA & 8.85 & -15.43 & -3.82 & 0.20 & 0.19 & 0.19 & -- \\
    NGC2764 & 330 & 39.60 & ATLAS3D & 8.44 & -15.69 & -3.26 & 0.12 & 0.24 & 0.19 & -- \\
    NGC6478 & 331 & 105.10 & CALIFA & 8.97 & -15.23 & -5.04 & 0.25 & 0.07 & 0.19 & -- \\
    NGC2410 & 332 & 70.70 & CALIFA & 8.78 & -15.37 & -4.17 & 0.21 & 0.15 & 0.19 & -- \\
    NGC2449 & 333 & 74.00 & CALIFA & 8.76 & -15.44 & -3.99 & 0.20 & 0.17 & 0.18 & -- \\
    UGC10337 & 334 & 138.60 & CALIFA & 9.00 & -15.31 & -4.60 & 0.23 & 0.11 & 0.18 & -- \\
    NGC2481 & 335 & 32.00 & ATLAS3D & 8.53 & -15.46 & -3.95 & 0.19 & 0.17 & 0.18 & -- \\
    NGC0214 & 336 & 64.90 & CALIFA & 8.70 & -15.47 & -3.93 & 0.19 & 0.18 & 0.18 & -- \\
    NGC6301 & 337 & 129.50 & CALIFA & 8.98 & -15.32 & -5.09 & 0.23 & 0.07 & 0.17 & -- \\
    NGC4047 & 338 & 56.70 & CALIFA & 8.51 & -15.73 & -3.55 & 0.11 & 0.21 & 0.17 & -- \\
    NGC0257 & 339 & 75.10 & CALIFA & 8.66 & -15.60 & -3.83 & 0.15 & 0.19 & 0.17 & -- \\
    MCG-02-03-015 & 340 & 82.60 & CALIFA & 8.43 & -16.03 & -3.29 & 0.02 & 0.24 & 0.17 & -- \\
    UGC10710 & 341 & 130.10 & CALIFA & 8.93 & -15.39 & -4.66 & 0.21 & 0.11 & 0.17 & -- \\
    NGC7466 & 342 & 109.70 & CALIFA & 8.60 & -15.87 & -3.42 & 0.07 & 0.22 & 0.16 & -- \\
    NGC3630 & 343 & 25.00 & ATLAS3D & 8.42 & -15.53 & -4.20 & 0.17 & 0.15 & 0.16 & -- \\
    NGC2916 & 344 & 59.60 & CALIFA & 8.60 & -15.60 & -4.15 & 0.15 & 0.16 & 0.15 & -- \\
    NGC5720 & 345 & 121.60 & CALIFA & 8.74 & -15.68 & -4.04 & 0.12 & 0.17 & 0.15 & -- \\
    UGC03151 & 346 & 62.60 & CALIFA & 8.61 & -15.61 & -4.32 & 0.15 & 0.14 & 0.14 & -- \\
    UGC04197 & 347 & 69.80 & CALIFA & 8.55 & -15.76 & -4.00 & 0.10 & 0.17 & 0.14 & -- \\
    NGC6394 & 348 & 130.00 & CALIFA & 8.80 & -15.60 & -4.80 & 0.15 & 0.10 & 0.12 & -- \\
    UGC03969 & 349 & 121.40 & CALIFA & 8.50 & -16.08 & -4.15 & 0.01 & 0.16 & 0.11 & -- \\
    UGC00005 & 350 & 105.00 & CALIFA & 8.71 & -15.66 & -4.91 & 0.13 & 0.09 & 0.11 & -- \\
    UGC00987 & 351 & 66.50 & CALIFA & 8.41 & -15.97 & -4.24 & 0.04 & 0.15 & 0.11 & -- \\
    NGC5930 & 352 & 46.20 & CALIFA & 8.44 & -15.76 & -4.83 & 0.10 & 0.09 & 0.10 & -- \\
    NGC7047 & 353 & 87.60 & CALIFA & 8.66 & -15.67 & -5.83 & 0.13 & 0.00 & 0.09 & -- \\
    NGC5980 & 354 & 68.90 & CALIFA & 8.56 & -15.74 & -5.18 & 0.11 & 0.06 & 0.09 & -- \\
    NGC0551 & 355 & 74.50 & CALIFA & 8.45 & -15.95 & -4.75 & 0.05 & 0.10 & 0.08 & -- \\
    IC1199 & 356 & 78.40 & CALIFA & 8.49 & -15.91 & -5.01 & 0.06 & 0.08 & 0.07 & -- \\
    NGC4185 & 357 & 64.90 & CALIFA & 8.49 & -15.82 & -5.34 & 0.08 & 0.04 & 0.07 & -- \\
    UGC12810 & 358 & 117.50 & CALIFA & 8.58 & -15.93 & -5.28 & 0.05 & 0.05 & 0.05 & -- \\
    MCG-02-51-004 & 359 & 87.60 & CALIFA & 8.50 & -15.93 & -5.44 & 0.05 & 0.04 & 0.04 & -- \\
\end{longtable}

\normalsize

\vspace{-10mm}

\bibliography{main}

\end{document}